\documentclass[aps,prx,twocolumn]{revtex4-1}

\usepackage{amsmath,amsfonts}
\usepackage{amsthm}

\usepackage{bbm,graphicx,bm,hyperref,xcolor}   

\makeatother
\def\ave#1{\langle #1 \rangle}
\def\ii{{\rm i}}
\def\sx{\sigma^{\rm x}}
\def\sy{\sigma^{\rm y}}
\def\sz{\sigma^{\rm z}}
\def\tr#1{{\rm tr}(#1)}
\def\1{\mathbbm{1}}

\def\ket#1{{| #1 \rangle}}
\def\kket#1{{| #1 \rangle\!\rangle}}

\def\bracket#1#2#3{{\langle #1 | #2| #3 \rangle}}
\def\bbracket#1#2#3{{\langle\!\langle #1 | #2| #3 \rangle\!\rangle}}
\def\bbraket#1#2{{\langle\!\langle #1 | #2 \rangle\!\rangle}}

\def\hx{h_{\rm x}}
\def\hz{h_{\rm z}}

\def\x0{\mathbf{x}_0}
\def\tH{t_{\rm H}}
\def\M{{\cal M}}
\def\S{{\cal S}}
\def\e{{\rm e}}
\def\T{{\cal T}}
\def\HF{H_{\rm F}}

\def\Aa#1{A^{(\boldsymbol{#1})}}
\def\a#1{a^{(\boldsymbol{#1})}}

\def\tit#1{{\em #1},}

\newcommand{\New}[1]{{#1}}
\newcommand{\new}[1]{{#1}}

\usepackage{booktabs}
\usepackage{multirow}
\AtBeginDocument{
\heavyrulewidth=.08em
\lightrulewidth=.05em
\cmidrulewidth=.03em
\belowrulesep=.65ex
\belowbottomsep=0pt
\aboverulesep=.4ex
\abovetopsep=0pt
\cmidrulesep=\doublerulesep
\cmidrulekern=.5em
\defaultaddspace=.5em
}

\begin{document}

\title{
    Prethermalization, shadowing breakdown, and the absence of Trotterization transition in quantum circuits
}

\author{Marko \v Znidari\v c}
\affiliation{Physics Department, Faculty of Mathematics and Physics, University of Ljubljana, 1000 Ljubljana, Slovenia}



\date{\today}

\begin{abstract}
  One of the premier utilities of present day noisy quantum computers is simulation of many-body quantum systems. We study how long in time is such a discrete-time simulation representative of a continuous time Hamiltonian evolution, namely, a finite time-step introduces so-called Trotterization errors. We demonstrate that the truncated operator propagator (Ruelle-Pollicott resonances) is a powerful tool to that end, as well as to study prethermalization and discrete time crystals, including finding those phenomena at large gate duration. We show that the effective energy is more stable than suggested by Trotter errors -- a manifestation of prethermalization -- while all other observables are not. Even the most stable observable though deteriorates in the thermodynamic limit. Different than in classical systems with the strongest chaos, where the faithfulness time (the shadowing time) can be infinite, in quantum many-body chaotic systems it is finite. A corollary of our results is also that, opposite to previous claims, there is no Trotterization transition in non-integrable many-body quantum systems. \new{We demonstrate our results on a one-dimensional (1d) kicked Ising model, as well as on 1d kicked XX model and 2d kicked Ising model. The truncated propagator is also used to calculate the energy diffusion constant in the tilted-field Ising model with high accuracy.}
\end{abstract}



\maketitle

\section{Introduction}

One of the defining property of chaotic systems~\cite{Ott} is exponential complexity of their time evolution as quantified by positive dynamical entropies. A manifestation of that is, for instance, the colloquial ``butterfly effect'' -- two close initial conditions will separate exponentially fast in time. Considering that typical chaotic systems are not analytically solvable, and one therefore has to resort to the indignity of numerical simulations, a pertinent question is how representative are such noisy numerical trajectories of true trajectories? At first sight it would seem that the exponential separation of trajectories renders any noisy numerical simulation futile. However, things are not as grim. 

First, we should specify what is it that we really want from a noisy simulation. Comparing a noisy and a true trajectory both starting from the same specific initial condition is not what we should in most cases care about. This is especially true in a many-body system, where we do not have access to all degrees of freedom, and in quantum mechanics where the stress is on (ensemble) averages rather than on individual pure states. For instance, calculating the overlap of two states, often called the fidelity (a.k.a. the Loschmidt echo), is not the most informative quantity in a many-body system (unless one is e.g. interested in the ground-state physics of a gapped system): small fidelity does not necessarily mean that few-body observables that we care about have gone awry. Furthermore, the properties of a specific exact initial state are really irrelevant; measurable quantities away from phase transitions should be robust to small variations. What one should care about is that the noisy evolution of a particular initial state is still {\em representative} of a true non-noisy evolution, i.e., that its results are similar to an exact average. It would be enough for the noisy evolution to be similar to a true evolution from some nearby initial condition. Precisely that property is captured by the famous shadowing property of chaotic trajectories~\cite{Ott} that comes to our rescue when doing noisy simulations of chaotic systems. 

\begin{figure}[t]
  \centerline{\includegraphics[width=0.35\textwidth]{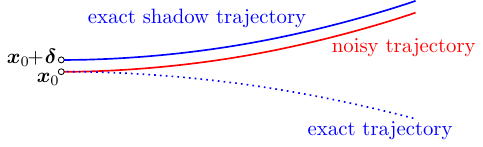}}
  \caption{{\bf Shadowing in classical chaotic systems.} A true shadow trajectory (blue) from a slightly perturbed initial condition closely follows (shadows) the noisy one (red), that otherwise exponentially separates from the original exact trajectory (dotted blue).}
  \label{fig1}
\end{figure}
Shadowing means that, while it is true that a trajectory $f^t_\epsilon(\x0)$ starting from $\x0$ will in the presence of noise $\epsilon$ diverge exponentially fast from a true one $f^t_0(\x0)$ starting from $\x0$, there exists a nearby initial condition $\x0'=\x0+\boldsymbol{\delta}$ for which a true trajectory $f^t_0(\x0')$ closely follows the original noisy one, $f^t_0(\x0') \approx f^t_\epsilon(\x0)$. \new{In other words, while a noisy trajectory is not a good representative of an exact one, it closely follows a ``shadow trajectory'' that starts from a slightly perturbed initial condition (Fig.~\ref{fig1}). The shadowing time is the time up-to which one can find the shadow trajectory that follows the noisy one.} For uniformly hyperbolic~\cite{foothyper} systems the above intuitive picture can be made rigorous in the form of the shadowing lemma~\cite{Anosov67,Bowen75}, \new{saying that the shadowing time in such systems is infinite. Uniformly hyperbolic systems are in a way the nicest chaotic systems (strongest chaos), an example being the Arnold's cat map, in which noisy trajectories are representative of exact ensemble averages.} For classical systems that are not uniformly hyperbolic the shadowing time is finite~\cite{Yorke97}.

We address the question of the shadowing time in quantum many body systems. This is an important theoretical question, that is moreover very relevant for noisy quantum computers and simulators, whose best-case present use~\cite{troyer23} is to simulate~\cite{daley22} other quantum systems~\cite{ibm,google,dario,shtanko25,feig25}, including the prethermalization regime~\cite{banuls23} that we shall discuss. If one wants to demonstrate the supremacy of quantum computation the number of qubits has to be large, and the simulation has to be faithful to true dynamics. Simulating the many-body quantum Hamiltonian $H_0=\sum_j h_j$ with local interactions $h_j$ can be implemented by splitting the time evolution propagator
\begin{equation}
  U(\T)=\exp{(-\ii H_0 \T)}
  \label{eq:U}
\end{equation}
into small timesteps of duration $\tau$, thereby approximating the Hamiltonian evolution by a quantum circuit,
\begin{equation}
  U(\T) \approx U^t_\tau,\quad U_\tau=\prod_j \exp{(-\ii h_j \tau)},\quad t=\frac{\T}{\tau}
  \label{eq:trott}
\end{equation}
where each term $\exp{(-\ii h_j \tau)}$ is now, for instance, a 2-qubit gate that can be implemented on a quantum computer. A natural question is how is with the shadowing property in such many-body quantum systems; how long can a given simulation (RHS of Eq.(\ref{eq:trott})) be trusted to represent the true dynamics (LHS of Eq.(\ref{eq:trott}))?

Characterizing shadowing properties of a many-body quantum system is a formidable task, \new{first because of all the intricacies that quantum mechanics brings, and second because of the system size limit $L\to \infty$, where even for classical systems things are not clear. We will therefore tackle a simpler question. We will not address the full shadowing, which is a statement on closeness of states, i.e., of all observables, but rather a question of direct experimental relevance that will nevertheless shed light also on the shadowing time.} We will (i) focus on a specific type of noise, namely errors~\cite{childs19,cubitt21,Heyl} due to Trotterizing Hamiltonian evolution (\ref{eq:trott}), and (ii) measure closeness by simple expectation values or autocorrelation functions of local observables (rather than some state norm).

\subsection{Summary of results}

\new{We will find that even in the optimal situation -- looking at the best-behaving observable -- a noisy simulation follows a true one only up-to a finite but long time, meaning that the shadowing time is necessarily finite. More specifically, under Trotterization errors there is a special observable,} namely, the energy $H_0$, that is very stable and resilient to errors, more than the simple Trotter error ${\cal O}(\tau^2)$ would suggest. Despite its stability though, with increasing system size $L$ the stability will deteriorate. \new{While in a finite system it might look that $H_0$ is stable, in the thermodynamic limit} of $L \to \infty$ and time $t \to \infty$, no matter how small $\tau$, the energy $H_0$ will not be conserved implying that the noisy simulation is fundamentally different than the true one. As a consequence, perfect infinite-time shadowing, like the one holding for classical systems with the strongest chaos, is impossible under Trotterization errors in locally interacting quantum many-body systems. The relevant timescale of the above instability is nothing but the prethermalization~\cite{Mori16,Abanin17} time $T$, for a review of prethermalization see Ref.~\cite{ho23}, and is given by the largest eigenvalue of the momentum-resolved truncated operator propagator. As predicted theoretically, it is exponentially large in the gate frequency $1/\tau$, but the benefit of our approach will be to give an exact quantitative prediction for $T$, not just a bound. The corresponding eigenvector is on the other hand an almost conserved effective energy which is for small $\tau$ close to $H_0$. \new{Note that the prethermalization time and the shadowing time are different concepts. The shadowing time is related to the worst-case stability -- thermalization of observable that deviates first, whereas we are focusing on the best-case scenario, i.e., thermalization of $H_0$. An alternative viewpoint is that we are interested in the breaking of a single conserved operator $H_0$ due to Trotterization errors.}

The truncated propagator therefore offers a way to obtain the effective Hamiltonian useful in studies of Floquet systems~\cite{Eckardt17}, and is an explicit and concrete alternative to various divergent operator expansions like the Baker-Cambell-Hausdorff (BCH) or the Floquet-Magnus series~\cite{blanes09}. We also point out and numerically demonstrate that in order to infer the correct behavior in the thermodynamic limit the Heisenberg time $\tH$ (the inverse level spacing) has to be much larger than any relevant dynamical timescale in the system, specifically the long prethermalization time $T$. \new{It is another example when relying on small-system results, for instance obtained by exact diagonalization, can lead to incorrect conclusions.} Previous claims~\cite{Heyl,Polkovnikov,Bukov} of a small Trotter step transition (or a crossover) in system's properties, like the scaling of errors, were a finite size effect caused by a finite averaging time (i.e., being in the incorrect limit of $T>\tH$). In the thermodynamic limit (TDL) there is no transition.

Another utility of the truncated propagator is to study discrete time crystals (DTC)~\cite{vedika16,else16,curt16}, see Ref.~\cite{zaletel23} for a review. Because we have access to exact $T$ we will be able to address the question of the stability and the thermodynamic limit behavior of the DTC in a nearest-neighbor kicked Ising model away from its integrable $\tau=\pi$ point. We find that the prethermalization, underlying the robustness of DTC, is rather prevalent in the kicked Ising model and not limited only to small values of $\tau$. Finally, looking at the quasimomentum dependence of the spectral gap we will be able to calculate the diffusion constant, either during a finite but long prethermalization time, or in the Hamiltonian limit. Because we directly work in the thermodynamic limit $L\to \infty$ the value of the energy diffusion constant for the chaotic Ising model is calculated with higher precision than with other methods. We also briefly show results for the kicked XX model and a two-dimensional (2D) kicked Ising model, demonstrating generality of our results.

\New{The truncated operator propagator method, even in our present brute-force implementation whose complexity grows exponentially with the truncated operator size $r$, offers a procedure that can treat difficult problems with long time-scales, like prethermaliation or diffusion, better than existing methods, obtaining quantitatively precise results in the thermodynamic limit with relatively modest numerical effort.}

\section{Observables and truncated operator propagator}

As our criterion of closeness of $U$ to its ``noisy'' approximation $U_\tau^t$ we shall look at how close are either expectation values of observables, or more generally, correlation functions of observables. Expectation values will be evaluated at infinite temperature,
\begin{equation}
  \ave{A} \equiv \tr{A}/2^L,
  \label{eq:A}
\end{equation}
where $2^L$ is the Hilbert space size of $L$ qubits. Compared to using averaging over a specific state, such an infinite-temperature average over density operator $\rho=\1/2^L$ is unbiased, easier to treat theoretically, and is also experimentally relevant. Due to self-averaging it can be replaced at large $L$ by an average over a single random state $\ket{\psi}$, $\ave{A} \approx \bracket{\psi}{A}{\psi}$. More informative than a simple expectation value is the autocorrelation function $C_A(t)$ of observable $A$,
\begin{equation}
    C_A(t)\equiv \frac{1}{L} \ave{A(0)A(t)},
  \label{eq:C}
\end{equation}
where we will be interested in $A$'s that are extensive sums of local densities,
\begin{equation}
  A=\sum_j a_j,
  \label{eq:a}
\end{equation}
such that the prefactor $1/L$ in Eq.(\ref{eq:C}) ensures that $C_A(t)$ becomes system size independent for large $L$. Concretely, we shall take for $A$ either the total magnetization $Z=\sum_j \sz_j$, the energy $H_0$, or the staggered magnetization $S=\sum_j (-1)^j \sz_j$.

\begin{figure*}[ht!]
  \centerline{\includegraphics[width=\textwidth]{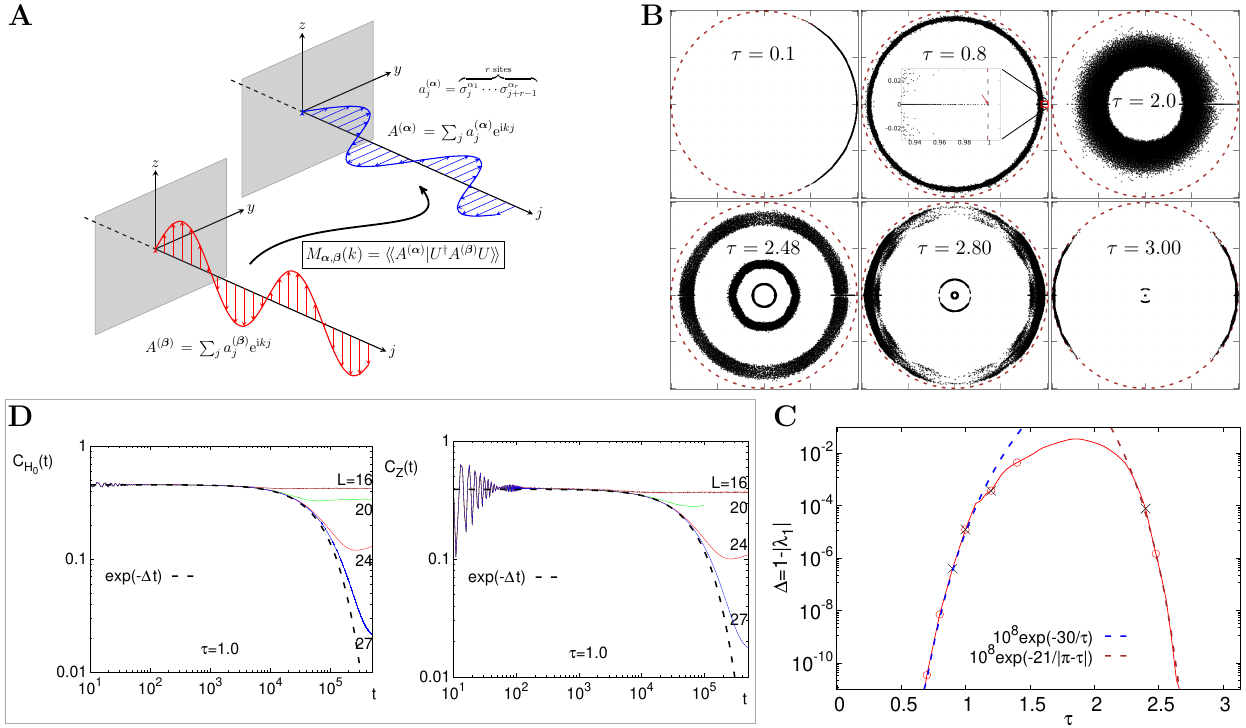}}
  \caption{{\bf Truncated propagator and Ruelle-Pollicott resonances.} ({\bf A}) Momentum-resolved operator propagator $M(k)$ is truncated to infinite system size operators with their density supported on at most $r$ sites \new{(sine-shaped curves suggest quasimomentum modulation of local densities, e.g., of $\a{{\alpha}}_j=\sz_{j}$ having $r=1$)}. ({\bf B}) Spectrum of non-unitary $M(k=0)$ lies inside a dashed unit circle. Shown are spectra at truncation $r=8$ and different gate duration $\tau$ in the kicked Ising model (\ref{eq:KIU}). Spectrum can be used to identify prethermalization (e.g. $\tau=0.8$) as well as a discrete time crystal ($\tau=3.0$). The largest eigenvalue (e.g., zoom-in inset with red arrow in the top middle frame) is shown in (C). ({\bf C}) Dependence of the spectral gap $\Delta$ on $\tau$ \New{(full curve is for $r=9$)}. Circles and crosses mark $\tau$ for which we show \New{in other figures} convergence with $r$, and correlation functions, respectively. ({\bf D}) Autocorrelation function of energy $H_0$ (\ref{eq:H0}) and magnetization $Z=\sum_j \sz_j$ (right). Dotted curve is exponential decay with the rate $\Delta$ from (C), showing that the inverse gap $T=1/\Delta$ gives the duration of the prethermalization plateau. In the thermodynamic limit $L \to \infty$ the plateau at large time disappears for any $\tau$ -- there is no Trotterization transition with $\tau$. \New{Data is for the kicked Ising model (\ref{eq:KIU}) with $J=\frac{1}{2}$, $\hx=\hz=1$.}}
\label{fig2}
\end{figure*}
To evaluate $C_A(t)$ we need the time evolved observable $A(t)=U_\tau^{-t} A U_\tau^t$, which can be written in the Hilbert space of operators as $\kket{A(t)}=\M^t \kket{A}$, where $\M$ is a linear (super)operator propagator that propagates observables (operators) by one step (one $U_\tau$). With this notation the correlation function is simply
\begin{equation}
  C_A(t)=\bbracket{A}{\M^t}{A},
  \label{eq:AMA}
\end{equation}
where the Hilbert-Schmidt inner product is $\bbraket{A}{B}\equiv \tr{A^\dagger B}/\tr{\1}$. The expectation value is on the other hand $\ave{A(t)}=\bbracket{\1}{\M^t}{A}$. In systems that have a mixing property (connected) correlation functions must decay to zero, while in chaotic ones without conserved operators one moreover expects this decay to be exponential for local $A$. One might wonder how can one obtain decaying correlation functions from the properties of $\M$? This is not obvious because $\M$ is unitary and therefore its spectrum is on a unit circle -- one would expect that e.g. exponential decay of $C_A(t)$ will be according to Eq.(\ref{eq:AMA}) connected to eigenvalues of $\M$ that satisfy $|\lambda|<1$. Unitarity of $\M$ however is not in contradiction with the apparent relaxation (equilibration): namely, we expect relaxation of local observables and related decay of correlation functions; under unitary evolution no information is lost, it just gets spread out into the full Hilbert space, i.e., ``disappears'' in non-local operators.

In order to recover the decay of correlations from $\M$ one must look at analytical properties of its resolvent $(\M-z\1)^{-1}$ inside a unit circle of complex $z$. This idea is known under the name of Ruelle-Pollicott (RP) resonances~\cite{Ruelle86,Pollicott85,gaspard}. One way to get access to possible singularities inside a unit circle is to limit oneself to an appropriate functional space -- in our case it will be the space of local operators.

\subsection{Truncated propagator}

The truncated operator propagator is the main technical tool of this work and will enable us to quantitatively study the prethermalization, DTC, and more, thereby resolving several outstanding questions.

One way to break the unitarity of $\M$ is to truncate it to operators with local support. Doing that one obtains a finite dimensional truncated $\M$ whose eigenvalues $\lambda_j$ move inside a unit circle~\cite{Prosen}. The largest one in modulus $\lambda_1$ will, provided it is isolated from the rest, determine the asymptotic decay of correlations as $C_A(t) \sim \lambda_1^t$, and is called a Ruelle-Pollicott resonance~\cite{Ruelle86,Pollicott85}, a resonance inside a unit circle in the analytical continuation of the propagator resolvent. In Ref.~\cite{Prosen} $\M$ was truncated to translationally invariant operators ($k=0$ in our notation) and studied in the kicked Ising model. Besides chaotic systems, the method can also be used to find constants of motion~\cite{marcin}, for instance in Refs.~\cite{katja,rustem} in cellular automata, and in Ref.~\cite{U1} to discover that all magnetization conserving brickwall qubit circuits are integrable.

We shall use an extension of the above method to an arbitrary (quasi)momentum $k$~\cite{RP24}, a refinement that has been already used to find new unconventional symmetries in integrable circuits~\cite{kpzwall}. Momentum dependence will allow us to treat any observable, not just translationally invariant ones, as well as deduce transport properties, for instance, calculate the value of the diffusion constant. Namely, we are interested in translationally invariant $H_0$ without any spatial inhomogeneities, such that also its Trotterization $U_\tau$ will have homogeneous gates. In other words, $U_\tau$ commutes with a shift by one site operator $\S$ (the method can be generalized to systems invariant only under a shift by more than one site~\cite{urban}), and therefore $\M$ is block diagonal in the eigenbasis of $\S$. Blocks can be labeled by eigenvalues $\e^{\ii k}$ of $\S$, where $k\in [-\pi,\pi]$ is a lattice quasimomentum (we shall often call it just momentum). The operator basis in a given momentum block consists of operators $\Aa{\alpha}$,
\begin{equation}
  \Aa{\alpha}=\sum_j \e^{\ii k j} \a{\alpha}_j,\quad \a{\alpha}_j=\sigma_j^{\alpha_1}\cdots \sigma_{j+r-1}^{\alpha_r},
  \label{eq:Pauli}
\end{equation}
where the sum runs over all site indices $j$ (in an infinite system), and $\sigma^{\alpha_j}\in \{\1,\sx,\sy,\sz \}$ are Pauli matrices (the local operator basis) specified by a vector of indices $\boldsymbol{\alpha}=(\alpha_1,\ldots,\alpha_r)$. Matrix elements of the operator propagator are (see Fig.~\ref{fig2}A for illustration)
\begin{equation}
  M_{\boldsymbol{\alpha},\boldsymbol{\beta}}(k)=\bbraket{\Aa{\alpha} }{U_\tau^{\dagger} \Aa{\beta} U_\tau }.
\label{eq:M}
\end{equation}
While it looks that in order to calculate $M(k)$ one has to deal with infinite-size operators $\Aa{\alpha}$, because of a sharp light-cone in quantum circuits the initial local density $\a{\alpha}_j$ can in one step spread over at most a fixed number $s$ of additional sites to the left or to the right of its initial support. For our specific kicked Ising circuit (Fig.~\ref{fig3}) where $U_\tau$ contains only a single layer of commuting nearest-neighbor 2-qubit gates we have $s=1$. It is therefore enough to work with propagation on just $r+2$ sites. \new{One can in fact implement evolution of local operators in-situ using only $r$ sites instead of $r+2$, see Appendix~\ref{app:insitu} for details, which enables us to study larger truncation sizes $r$, or simply speeding up calculations compared to tracking evolution on full $r+2$ sites.} Denoting $b(\boldsymbol{\beta}) \equiv U_\tau^\dagger \a{{\beta}}_1 U_\tau$, Eq.(\ref{eq:M}) explicitly evaluates to
\begin{equation}
  M_{\boldsymbol{\alpha},\boldsymbol{\beta}}(k)= \frac{1}{2^{r+2}}{\rm tr}\left[ \left( \a{\alpha}_1 + \a{\alpha}_0 \e^{\ii k} + \a{\alpha}_2 \e^{-\ii k}\right) b(\boldsymbol{\beta}) \right].
  \label{eq:Mk}
\end{equation}
Such an operator propagator has a symmetry $M^\dagger(U)=M(U^\dagger)$ and is without truncation a unitary operator (for $k=0$ and Hermitian basis, like the Pauli basis (\ref{eq:Pauli}), it is real orthogonal). If we truncate the basis to all local densities supported on at most $r$ consecutive sites (1d systems), that is vector indices $\boldsymbol{\alpha}$ and $\boldsymbol{\beta}$ run over $N=3\times 4^{r-1}$ possible local densities ($3$ instead of $4$ comes because the first index $\alpha_1$ should not describe $\1$, otherwise the density would be supported on at most $r-1$ sites, \new{and could be obtained by a translation of other basis elements}), we get a truncated operator propagator that is the main workhorse of this work. Due to the truncation we are dropping some matrix elements of an infinite-size unitary operator, resulting in a finite non-unitary matrix of size $N \times N$. For more details on implementation see Refs.~\cite{RP24,urban}\new{, and for in-situ propagation Appendix~\ref{app:insitu}.}

\subsection{Kicked Ising model}
\label{sec:KI}

Properties of $M(k)$ (\ref{eq:Mk}) will allow us to answer all the questions that we highlighted in the introduction. Let us demonstrate its spectral properties on the tilted field Ising model, which is a canonical many-body system that is chaotic for an appropriate parameter choice, e.g. $\hx=\hz=1$~\cite{pre07}.

\begin{figure}[t!]
  \includegraphics[width=0.22\textwidth]{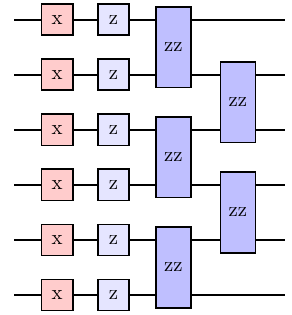}
  \caption{{\bf Kicked Ising quantum circuit.} Sequence of gates for one Floquet step of the kicked Ising model (\ref{eq:KIU}).}
  \label{fig3}
\end{figure}

The Hamiltonian is
\begin{equation}
  H_0=J\sum_j \frac{\sz_j \sz_{j+1}}{2}+\hz \sz_j+ \hx \sx_j.
  \label{eq:H0}
\end{equation}
Evolution $U(\T)$ can be implemented on a quantum computer by splitting $H_0$ into two terms, one containing $\sz$ operators and the other $\sx$. The one-step propagator shown in Fig.~\ref{fig3} is,
\begin{eqnarray}
  \label{eq:KIU}
  U_\tau&\equiv& U_{\rm z} U_{\rm x},\\
  U_{\rm x}&=&\prod_j \e^{-\ii \tau \hx \sx_j/2},\nonumber \\
  U_{\rm z}&=&\e^{-\ii \tau \sum_j \sz_j \sz_{j+1}/4}\prod_j \e^{-\ii \tau \hz \sz_j/2}. \nonumber
\end{eqnarray}
Alternatively, the same $U_\tau$ can be thought of as being a propagator of a time-periodic Hamiltonian,
\begin{equation}
  H=J\sum_j \left[ \frac{\sz_j \sz_{j+1}}{2}+\hz \sz_j+ \hx \sx_j\!\!\sum_{n=-\infty}^{\infty}\!\!\tau \delta(t-n\tau) \right],
  \label{eq:KI}
\end{equation}
called the kicked Ising model. RP resonances in the truncated propagator of the kicked Ising model have been studied in Refs.~\cite{Prosen,prosen07}, and more recently in Ref.~\cite{RP24}.

There are three parameters $\tau$, $\hx$, and $\hz$. For generic parameters the model is nonintegrable without any local conserved quantities. We measure time in units of $\tau$, i.e., our $t$ takes integer values, and we use parameters $J=\frac{1}{2}, \hx=\hz=1$, the same as e.g. Refs.~\cite{Heyl,Polkovnikov}. Only in section \ref{sec:DH} on transport will we use different ones in order to compare with values of the diffusion constant \New{at parameters} found in the literature.

\subsection{Spectrum of the truncated propagator}

Let us briefly describe how the spectrum of $M(k)$ (\ref{eq:Mk}) looks like -- in later sections we will describe in detail how it can serve as a useful tool to study Floquet systems (and also Hamiltonian ones in an appropriate limit).

In Fig.~\ref{fig2}B we show a set of exemplary spectra of $M(k)$ for $k=0$ and the truncation size $r=8$. Because we have $\hx=\hz=1$ the spectrum $\sigma(M,\tau)$ of $M(k)$ has a symmetry $\sigma(M,\tau)=\sigma(M,2\pi-\tau)$ (i.e., is symmetric around $\tau=\pi$, $\sigma(M,\pi+\tau)=\sigma(M,\pi-\tau)$), meaning that it is enough to study parameter range $\tau \in [0,\pi)$. In all cases the spectrum is within a unit circle, its form though heavily depends on the gate duration $\tau$. At very small $\tau=0.1$ the propagator $U_\tau$ is close to the identity and therefore all eigenvalues of $M(k)$ are close to the unit circle (within a band of width $1-|\lambda_j| \sim \tau^2$). At small~\cite{foot1} $\tau=0.8$, which will be a regime of most interest to us, one has a ring of eigenvalues, and with those closest to the unit circle lying on the real axis. Of particular interest is the largest one $|\lambda_1|$ because it will determine the asymptotic exponential decay rate of correlation functions (the inset in Fig.~\ref{fig2}B).

At $\tau>2$ the spectrum qualitatively changes and is made out of 3 rings. Point $\tau=\pi$ is special because it is integrable (two otherwise different integrable points merge into one because of $\hx=\hz=1$)~\cite{RP24,prosen07}: singular values of $M(k)$ are all $1$ ($U_{\rm x}=\prod_j (-\ii \sx_j), U_{\rm z}=\prod_j (-\ii \sz_j)\prod_p(1-\ii \sz_p \sz_{p+1})/\sqrt{2}$), while eigenvalues of $M(k)$ are $10\cdot 4^{r-3}$ equal to $+1$, the same number is $-1$, while the rest is $0$ (i.e., a fraction $5/24$ of all is $+1$). The fact that nonzero eigenvalues for $\tau \to \pi$ bunch around $\pm 1$ is a signature of a discrete time crystal~\cite{vedika16,else16,curt16}, specifically of the model discussed in e.g. Ref.~\cite{se21}. As we shall see, expectation values of some observables will exhibit oscillations with period $2$ (Sec.\ref{sec:DTC}). Spectrum of $M(k)$ will therefore tell us a lot of interesting things.

\section{Results}

\subsection{Leading Ruelle-Pollicott resonance and prethermalization}

The most important quantity is the eigenvalue of $M(k)$ with the largest modulus. It determines the long-time decay rate of correlation functions of generic observables (those not orthogonal to the corresponding eigenvector) under evolution by circuit $U_\tau$. Furthermore, as we shall see, it also determines the duration of the prethermalization phase, as well as a range of times and length-scales on which one can observe transient diffusion during the prethermalization phase, being a Floquet remnant of energy diffusion in the Hamiltonian $H_0$. Other methods have been used to find ``slowest'' observables, like minimizing the commutator with $U_\tau$~\cite{mari15} (fastest decay at short times), see also Refs.~\cite{motrunich17,izotova23}, or using time averaging~\cite{marcin15}.

In Fig.~\ref{fig2}C we show numerical results for the largest eigenvalue of $M(k)$, which turns out to be from $k=0$ (due to reflection symmetry there are two parity subsectors at $k=0$, and the shown $\lambda_1$ is from the even subsector, see also Ref.~\cite{RP24} for more details on parity sectors). We can see that the gap (distance to the unit circle) is rather small, and $\lambda_1$ is always isolated from the 2nd largest eigenvalue (will be shown later). We have checked that within the shown precision $\lambda_1$ is converged with respect to increasing the truncation support $r$ (see Appendix \ref{app:Num}). We shall especially focus on small $\tau$ behavior where the gap decreases rapidly with increasing gate frequency $1/\tau$, being a manifestation of prethermalization~\cite{Abanin17,Mori16} in which one has a parametrically long regime in which the system looks like being thermal with respect to an almost conserved Hamiltonian, see Ref.~\cite{ho23} for a review. While one can prove generic bounds on the prethermalization time $T$~\cite{Mori16,Abanin17,kuwahara16,abaninCMP}, they do not give an explicit prediction that could be used in a concrete model. The truncated operator propagator provides that.

For small $\tau$ we can try to write the propagator $U_\tau$ in terms of the Floquet Hamiltonian $H_{\rm F}$,
\begin{equation}
  U_\tau = \e^{-\ii \HF \tau} \approx \e^{-\ii \HF'\tau}.
  \label{eq:HF}
\end{equation}
The exact Floquet Hamiltonian $\HF$ is in general nonlocal and can not be obtained by e.g. BCH expansion because the series is divergent, see Appendix \ref{app:BCH}. However, up-to prethermalization time $T$ we can write the propagator approximately in terms of an almost conserved operator $\HF'$. While one can in principle obtain $\HF'$ via a BCH expansion, such a procedure is fraught with a certain freedom of what is the ``optimal'' truncation of an otherwise divergent series. The truncated operator propagator provides a more direct way -- $\HF'$ is nothing but the eigenvector corresponding to $\lambda_1$.

One expects that the prethermalization duration $T$ is exponentially large in $1/\tau$~\cite{Mori16,Abanin17,kuwahara16,abaninCMP}. A hand-waving argument is that if the frequency $1/\tau$ is larger than the local energy density $h$, it will take of order $\sim 1/(h\tau)$ simultaneous spin ``flips'' in order to absorb the energy $1/\tau$. Therefore, such a process will only occur at a high order $n \sim 1/\tau$ in the BCH expansion of $H_{\rm F}$, leading to an exponentially small probability, and as a consequence to an exponentially long time $T$ when the heating begins,
\begin{equation}
  T \sim \e^{a/\tau}.
\end{equation}
The truncated operator propagator gives, importantly, exact quantitative prediction, namely
\begin{equation}
  T=1/\Delta,\qquad \Delta \equiv 1-|\lambda_1|,
  \label{eq:T}
\end{equation}
where $\lambda_1$ is the largest eigenvalue of $M(k)$. For the chosen parameters the obtained numerical dependence can be described well by $T \approx 10^{-8} \exp{(30/\tau)}$, see Fig.~\ref{fig2}C (in the range $T \in [10^4,10^{12}]$; for smaller $\tau$ the gap gets too small for double precision arithmetic, while we could not reach convergence at available $r$ using higher precision arithmetic). The corresponding right eigenvector is equal to an almost conserved operator $\HF'$ (that could be approximately calculated in principle using the BCH series, see Ref.~\cite{RP24} for demonstration).

That $T$ is exactly the correct duration of the prethermalization phase is shown by calculating autocorrelation of various observables numerically exactly in finite systems (see Appendix \ref{app:Num} for numerical details). We specifically focus on the energy $H_0$ (\ref{eq:H0}) and the total magnetization $Z$. Both have a large overlap with the eigenvector corresponding to $\lambda_1$, and should therefore decay very slowly with time $t$. This is visible in Fig.~\ref{fig2}D. The theoretical dashed line shown in plots is exponential decay,
\begin{equation}
  C_A(t) \asymp B_A \e^{-\Delta t},
\end{equation}
without any fitting parameters. Its decay rate is $\Delta$ read off from Fig.~\ref{fig2}C, while its amplitude $B_A$ is calculated exactly to few lowest orders in $\tau$ using the BCH expansion, obtaining (for $J=1/2, \hx=\hz=1$)
\begin{equation}
  B_{H_0}=\frac{9}{16}-\frac{3}{32}\tau^2+\cdots,\qquad B_Z=\frac{4}{9}-\frac{4}{81}\tau^2+\cdots,
\end{equation}
see Appendix \ref{app:BCH} for details. We can see that this theory perfectly agrees with numerics, including at other values of $\tau$ (see Appendix \ref{app:Num} for data).

Important for our discussion of shadowing is behavior after prethermalization ends, $t \gg T$. We can see in Fig.~\ref{fig2}D that at finite $L$ a 2nd lower plateau is reached, which however decreases to $0$ as on increases $L$. In the thermodynamic limit, where one has to first send $L \to \infty$ and only then $t \to \infty$, there is therefore no plateau in correlation functions, nor in the magnetization itself. This is different than what has been observed in Refs.~\cite{Heyl,Polkovnikov,Bukov} -- different conclusions there were a consequence of using a finite averaging time over which one calculated e.g. magnetization (i.e., the averaging time was smaller than the prethermalization time $T$ (\ref{eq:T}), or, using too small $L$ for the chosen averaging time). In the TDL of chaotic systems there is therefore no Trotterization transition due to mysterious ``quantum localization''. What was observed in Ref.~\cite{Heyl} is a simple manifestation of a single almost conserved operator due to prethermalization. Let us remark that the Trotterization transition in integrable systems~\cite{vernier23}, for instance in XXZ-type integrable circuits, is of a completely different origin -- it is a consequence of different phases between which the whole analytical integrability structure changes~\cite{U1} and thereby also physical properties~\cite{kpzwall}.

If the shadowing time for quantum many-body systems would be infinite then the energy $H_0$ should be stable, i.e., there should exist a shadowing trajectory under noisy $U_\tau$ for which the expectation value of $H_0$ would stay close to the exactly constant value obtained under Hamiltonian evolution by $U$. Because this is not the case (Fig.~\ref{fig2}D) we can conclude that the shadowing in quantum many-body systems under Trotter errors can not be infinite. This is, in fact, in-line with expectations, classical and quantum. In quantum Floquet systems one can argue that eventually the system should ``heat-up'' (e.g. Refs.~\cite{rigol14,lazarides14}). In classical systems one for instance also finds that the shadowing time is finite, and in particular gets smaller if there is a large variation in the values of Lyapunov exponents~\cite{Yorke97,Grebogi90}. Many-body systems, where in the classical case one has a whole spectrum of Lyapunov exponents, have in a way too much ``freedom'' to be uniformly hyperbolic, and therefore have an infinite shadowing time.
\begin{figure}[t!]
   \centerline{\includegraphics[width=0.4\textwidth]{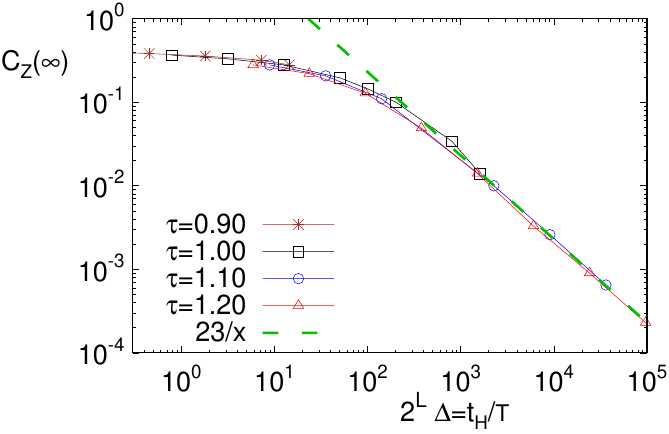}}
   \caption{{\bf Prethermalization vs. Heisenberg time.} Finite-size asymptotic value of magnetization correlation function (e.g., from Fig.~\ref{fig2}C we can read $C_Z(\infty) \approx 0.1$ for $L=24$ and $\tau=1.0$, reached at $t>10^5$) is shown for various sizes $L=14-28$ and $\tau$. To observe the correct thermodynamic limit $\lim_{t \to \infty} \lim_{L \to \infty}$, in which $C_Z(\infty)$ decays as $\propto 1/2^L$ (green line), one must have $T \ll t_{\rm H}$ -- the prethermalization time $T=1/\Delta$ has to be smaller than the Heisenberg time $t_{\rm H} \sim 2^L$. \New{Data is for the kicked Ising model (\ref{eq:KIU}) with $J=\frac{1}{2}$, $\hx=\hz=1$.}}
   \label{fig4}
\end{figure}
\begin{figure*}[ht!]
    \centerline{\includegraphics[width=0.9\textwidth]{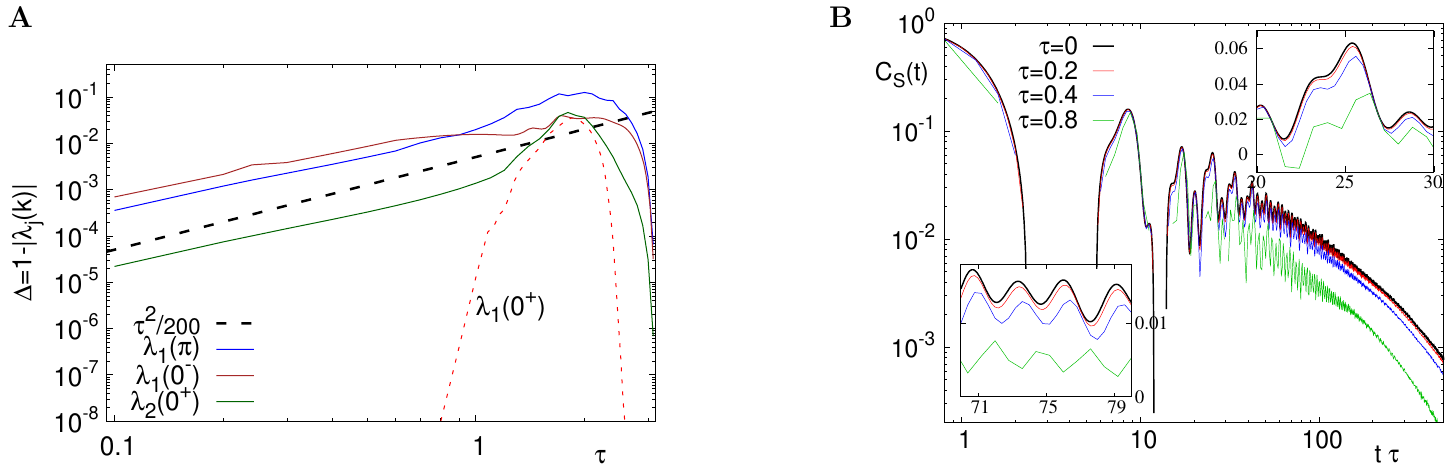}}
    \caption{{\bf Single almost-conserved operator.} ({\bf A}) Largest eigenvalues of $M(k)$ for different quasimomenta $k$: even and odd sector in $k=0$, and $k=\pi$. Only one gap is very small (dashed red, also Fig.~\ref{fig2}C), all other scale as $\sim \tau^2$ for small $\tau$. ({\bf B}) Correlation function of staggered magnetization $S=\sum_j (-1)^j \sz_j$ deteriorates fast already for small $\tau$ ($L=32$, finite-size effects are negligible on shown times; black curve is Hamiltonian evolution by $H_0$). The two insets zoom-in on specific time windows.}
\label{fig5}
\end{figure*}

\subsection{Finite-size effects and the Heisenberg time}
\label{sec:Heis}
From data like the one in Fig.~\ref{fig2}D we can see that at smaller $L$, like $L=16$ or $20$, even at large time autocorrelation functions do not decay but rather reach a 2nd nonzero plateau value $C_A(\infty)$. Why is that; how large $L$ do we need in order to observe the correct behavior in the TDL where correlation functions do decay? In a finite system the longest possible timescale is called the Heisenberg time and is given by the inverse level spacing. In a many-body system it has a leading exponential dependence on $L$, $t_{\rm H} \sim 2^L$. At larger times $t \gg t_{\rm H}$ quantum system resolves its discrete spectrum and essentially starts to behave like it is ``integrable'', even if its true dynamics is chaotic (e.g., one has recurrences, correlations do not decay, etc., due to the spectral resolution being a sum of a finite number of frequencies). One should therefore never look at properties of quantum systems at $t \gg t_{\rm H}$ -- at best one will sample finite-size effects.

Being careful about $t_{\rm H}$ is especially important in systems with long time scales, for instance, in our case the prethermalization time $T$ is exponentially large in $1/\tau$ (e.g., Fig.~\ref{fig2}C), another case is systems with disorder where dynamics gets exponentially slow with increasing disorder strength~\cite{panda19}. For instance, in Fig.~\ref{fig2}D we can see that at $L=16$ it indeed seems that even at $t\sim 10^6$ correlations do not decay, and one would be tempted to conclude that one is in a non-mixing regime. However, that is not correct. For larger $L$ correlations do start to decay. The reason for non-decay at $L=16$ is simply that one is not in the regime $T \ll t_{\rm H}$. \new{This is another example where relying on an exact diagonalization, either full spectrum or just an often-used diagonal approximation, would completely incorrectly diagnose the long-time dynamics in the thermodynamic limit (e.g., plateau in $L=16$ data in Fig.~\ref{fig2}D).}

In order to see the correct physics in the thermodynamic limit one should have
\begin{equation}
  T=\frac{1}{\Delta} \ll t_{\rm H} \sim 2^L.
  \label{eq:th}
\end{equation}
We numerically demonstrate correctness of Eq.(\ref{eq:th}) in Fig.~\ref{fig4}. From numerical simulation data we read out the value of $C_Z(t)$ at large time, where a 2nd plateau is reached for finite $L$. Plotting this value $C_Z(t \to \infty)$ as a function of scaled time $t_{\rm H}/T$, where $T$ is read off from Fig.~\ref{fig2}C, we get a good collapse of data for all $L$ and $\tau$ (i.e., $T$). We indeed see that in order to really see decaying correlations, where $C_Z(\infty) \sim 1/2^L$, one must have $t_{\rm H} \gg T$. Different conclusions in Refs.~\cite{Heyl,Polkovnikov,Bukov} are due to violation of this condition. Regarding the stability of expectation of $H_0$, the good news is that the prethermalization time is exponentially large in $1/\tau$ and so one does not need excessively small time step $\tau$ to observe long stability, however, if one wants to study the TDL one should still respect $T \ll \tH$ and therefore also not take a too small $\tau$ at a given $L$; the minimal meaningful timestep $\tau$ scales in a finite system as $1/L$ (\ref{eq:th}).

\New{While computational complexity of finding the largest eigenvalue of $M(k)$ in our implementation (see Appendix~\ref{app:Num}) grows exponentially with the support $r$, i.e., the necessary CPU time increases by $\approx 4$ when $r$ is increased by $1$, it is still much faster than any other known method used for prethermalization. For instance, taking as an example the kicked Ising at $\tau=1.0$ in Fig.~\ref{fig2}, exact pure-state evolution used to get the magnetization correlation function for $L=27$ until $t=5\cdot 10^5$ took $660$h ($\approx 27$ days on a single node of a standard computational cluster), while our calculation of the gap $\Delta$ at $r=10$ took $\approx 6$ minutes with the in-situ propagation (while it takes $\approx 9$h using slower propagation on $r+2$ sites, see Appendix~\ref{app:Num} for details). At $r=10$ the result $\Delta =1.20\cdot 10^{-5}$ is essentially converged with respect to $r$ (see Appendix~\ref{app:Num}), and even at much smaller truncation $r=6$, where the calculation takes only few seconds, we already get an almost correct $\Delta=1.33\cdot 10^{-5}$. On the other hand, for pure-state evolution decreasing $L$ is ``dangerous'': $L=24$ is barely enough (Fig.~\ref{fig2}D) to give us a hint of the correct prethermalization behavior (as discussed, this is the reason for previous incorrect claims) but still takes $\approx 4$ days. Using exact diagonalization, which is only possible at even smaller systems like $L=16$, would on the other hand give a completely incorrect physics at long times. In short, the truncated propagator method gives correct, high-precision results with much less numerical effort than other methods, while even at lower precision $r$, as opposed to other methods, it still identifies the correct physics (small $\Delta$ and prethermalization).}

\subsection{Fast decay and no prethermalization of generic observables}

Here we want to answer the question which observables do exhibit prethermalization? With a momentum-resolved propagator, which covers propagation of any observable, that is easy. We just have to check how many eigenvalues of $M(k)$ are exponentially close to a unit circle. As we shall see, there is only one.

Let us check if there are any other very small gaps of $M(k)$ which would imply that there is another almost conserved local operator. Numerically calculating the 2nd largest eigenvalue in the $k=0$ sector, or more precisely, the 2nd one in the even $k=0$ and the largest in the odd $k=0$ sector, we see in Fig.~\ref{fig5}A that those eigenvalues are always separated from the prethermalization one $\lambda_1(k=0)$. A similar situation occurs also for nonzero $k$. The spectral gap for all those eigenvalues scales as $\sim \tau^2$ for small timesteps, which is the same as the scaling of the Trotterization error ${\cal O}(\tau^2)$. We can therefore conclude that $\HF'$ is the only almost conserved local operator for which the prethermalization occurs.

\begin{figure*}[t!]
   \centerline{\includegraphics[width=0.9\textwidth]{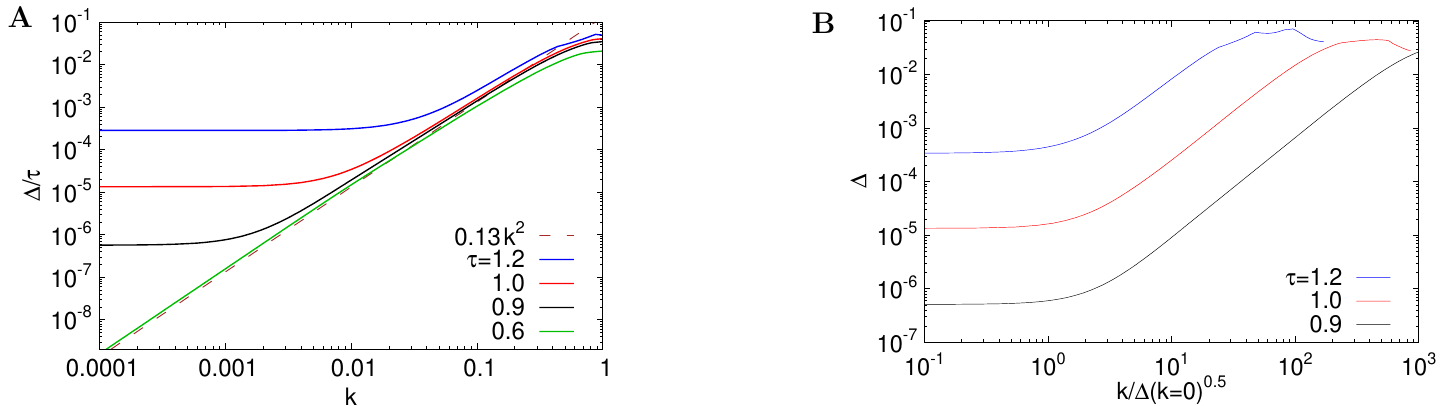}}
    \caption{{\bf Diffusion during prethermalization.} ({\bf A}) Dependence of the gap $\Delta=1-|\lambda_1(k)|$ on quasimomentum $k$ (at $k=0$ one gets Fig.~\ref{fig1}C, $r=6$). For small $k>k_0$ one has diffusive quadratic dependence $\Delta \approx D(\tau)\tau k^2$. ({\bf B}) Scaled $k$: diffusion can be observed at quasimomenta larger than $k_0 \sim \Delta^{1/2}$. The prefactor $D(\tau \to 0)$ is the diffusion constant of Hamiltonian $H_0$. \New{Kicked Ising model (\ref{eq:KIU}) with $J=\frac{1}{2}$, $\hx=\hz=1$.}}
\label{fig6}
\end{figure*}
All other observables that are orthogonal to $\HF'$ will deteriorate on timescales that are inline with Trotter errors, that is, much faster than the prethermalization time $T$. Note that while a typical observable from the even zero momentum sector will have a nonzero overlap with $\HF'$ (which though, depending on what we mean by typical, i.e. the measure, can be small), any observable with a nonzero $k$ will have zero overlap. Therefore, allowing observables with nonzero momentum a typical observable will not exhibit prethermalization. This is numerically demonstrated in Fig.~\ref{fig5}B. We pick a staggered magnetization $S \equiv \sum_j (-1)^j \sz_j$ and observe that the correlation function for finite $\tau$ starts to differ from the one obtained for Hamiltonian evolution generated by $H_0$ much more rapidly than e.g. $H_0$ or $Z$ in Fig.~\ref{fig2}D which do exhibit prethermalization. For the largest $\tau=0.8$ shown, for which the staggered magnetization visibly deviates already at $t \approx 22$, the total magnetization $Z$ would follow the Hamiltonian result up-to very large prethermalization $T \approx 10^8$. A typical observable is therefore much more sensitive to Trotterization errors because it does not exhibit prethermalization.

\subsection{Diffusion constant from the truncated propagator}

\subsubsection{Finite \New{gate} times}
Up-to large time $T$ the operator $\HF'$ behaves as if it would be conserved. Similarly as one can study energy transport under Hamiltonian evolution by $H_0$, one can also ask about transport properties of $\HF'$ in the circuit $U_\tau$ for $t \ll T$ when $\HF'$ is quasiconserved.

Assuming $H_0$ is chaotic, and thereby also $\HF'$, we would expect this transport to be diffusive. This is indeed the case and can be seen in the momentum dependence of the spectrum of $M(k)$. Details about the momentum-dependent spectrum of the truncated propagator in the presence of diffusion due to U(1) symmetry will be published elsewhere~\cite{urban}, for discussion of correlation functions see also recent Ref.~\cite{sarang25}, here we just look at $\lambda_1(k)$. Namely, the leading eigenvalue of $M(k)$ gives the asymptotic decay rate of correlation functions of operators with finite $k$. In particular, diffusion should be reflected in the decay rate of appropriate density correlations being $Dk^2$. Such a scaling can be obtained by Fourier transforming diffusion equation, or more explicitly, the correlation function of the diffusive conserved density $q$ in momentum space decays as $\ave{q(t,k)q(0,k)} \sim \e^{-Dk^2}$~\cite{Forster}.
\begin{figure*}[tb!]
      \centerline{\includegraphics[width=0.98\textwidth]{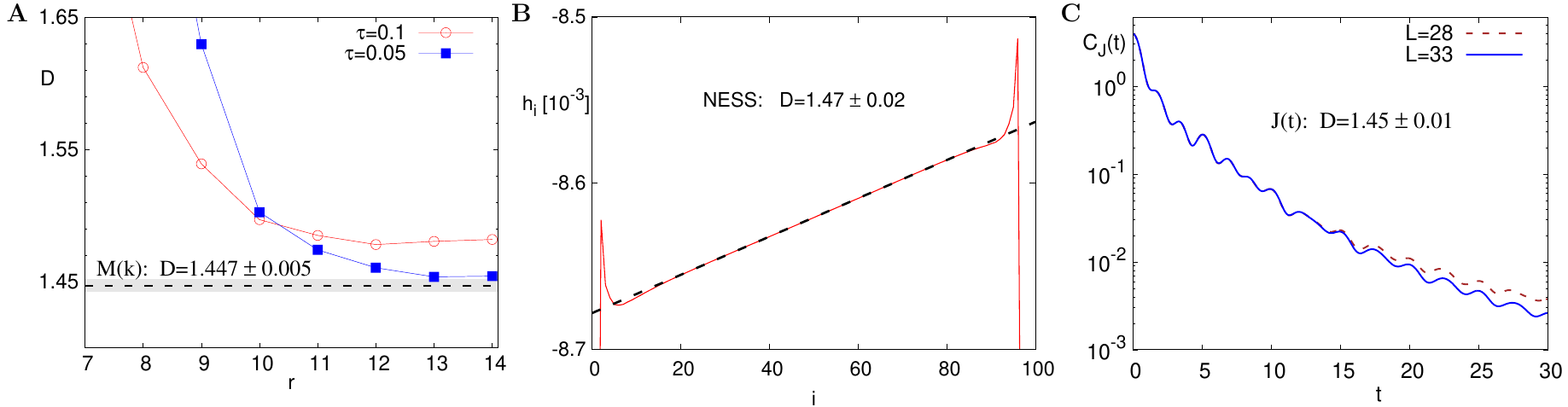}}
      \caption{{\bf Diffusion constant for the Hamiltonian tilted Ising model.} (A) Using the truncated propagator in the limit of small $\tau$ and $k$, getting $D=[1-\lambda_1(k)]/(k^2\tau)$ at $k=0.02$. (B) Lindblad NESS with $L=100$ and high-temperature boundary parameters. Shown is the energy density profile (the NESS current $j_p$ is independent of site $p$). (C) Green-Kubo and the energy current autocorrelation function. All three methods give consistently the same value of $D$ (${\cal J}=1.0, \hx=1.4, \hz=0.9045$).}
      \label{fig7}
\end{figure*}

We therefore have that the largest eigenvalue of the truncated operator propagator $M$ at $k=0$ determines the prethermalization duration $T=1/\Delta=1/(1-|\lambda_1(k=0)|)$, while its momentum dependence $1-\lambda_1(k) \approx Dk^2$ determines diffusion constant $D$. This can be seen in Fig.~\ref{fig6}\New{, from which we could for instance read out that a finite-time $D$ (until prethermalization ends) at $\tau=0.6$ is $D \approx 0.13\cdot 0.6=0.078$ (this is diffusion constant in units of time where one unit of time equals one propagation step of duration $\tau$; per unit of physical time, i.e., where one makes $1/\tau$ propagation steps in a unit of time, its value would instead be $D \approx 0.13$)}. We can in fact see that (i) one has quadratic dependence only at small $k$, i.e., diffusion is expectedly exhibited only on large hydrodynamic scales, and that (ii) $k$ must also not be too small in order to see diffusive $\Delta \sim k^2$. Looking at data for different $\tau$ we see that the $k_0$ above which one sees diffusive scaling goes as $\sqrt{\Delta}$ (Fig.~\ref{fig6}B). We remark that we do not find any subdiffusively scaling eigenvalues, which is different than what is found by minimizing the commutator~\cite{mari15}, i.e., in the short-time behavior.

\subsubsection{Hamiltonian limit}
\label{sec:DH}

For small $\tau$ the prefactor in front of $\sim k^2$ is trivially proportional to $\tau$ because in the Hamiltonian limit of $\tau \to 0$ one should recover the diffusion constant of the Hamiltonian $H_0$ which is measured in units of physical time $t\tau$, and not in time steps $t$ as obtained from the gap of $U_\tau$. To check that $D$ obtained in such a way from $M(k)$ is indeed equal to the diffusion constant of $H_0$ we have calculated $D$ directly for the Hamiltonian by two other methods, one is the Lindblad equation with boundary driving where one studies a true nonequilibrium steady state (NESS), while the other is the Green-Kubo formula using the equilibrium current autocorrelation function.

In order to facilitate comparison with existing data we here take slightly different parameters, namely, we use $H_0'=\sum_j {\cal J} \sz_j\sz_{j+1}+\hz \sz_j+\hx \sx_j$, with ${\cal J}=1$ and $\hx=1.4, \hz=0.9045$. \new{A number of recent papers has values of $D$ at infinite temperature for those parameters, with values in the range $D=1.38-1.55$, specifically, Ref.~\cite{tibor22} gets $D \approx 1.40$, Ref.~\cite{bard24} $D\approx 1.55$, Krylov method of Ref.~\cite{robin24} $D\approx 1.38$, with the most accurate value seemingly in Ref.~\cite{white24} at $D \approx 1.445$.

  Our results for $D$ obtained from a small $k$ behavior of $\lambda_1(k)$ of the truncated propagator are in Fig.~\ref{fig7}A. Taking small $\tau=0.05$ we get $D=(1-\lambda_1(k=0.02))/(k^2\tau)\approx 1.454$ for truncation size $r=13$, while doing extrapolation~\cite{footconv} to $r \to \infty$ gets us $D \approx 1.447 \pm 0.005$ (similar value is obtained for $k=0.01$). We see that our estimate agrees well with Ref.~\cite{white24}, while other methods are apparently less accurate. For larger $\tau=0.1$, when the Trotterization is not negligible, the converged value is a bit larger and reflects a finite-time diffusion constant in a circuit with long prethermalization time.}

We shall now compare the above value of $D$ with the one obtained by two additional completely different methods that study how the energy current density $j_{p}$ behaves,
\begin{equation}
    j_{p}=\hx{\cal J}( \sy_{p} \sz_{p+1}-\sz_{p-1} \sy_{p}).
  \label{eq:J}
\end{equation}
The first method relies on the scaling of energy current density $j_{p}$ in a nonequilibrium steady state (NESS), where according to Fourier's law one should have for NESS expectations $j_{p}=-D \nabla h$, with $h$ being the energy density and all quantities mean expectation value in the NESS. To induce a NESS we couple two spins at each chain end to a bath described effectively by 2-site Lindblad operators. We essentially repeat calculation from Ref.~\cite{jstat09} (see Appendix~\ref{app:Num}) for the parameters used here. Results are shown in Fig.~\ref{fig7}B. We also observed (data not shown) that as one lowers the temperature $D$ increases (e.g., decreasing the energy density to $h \approx -0.02$ we get about $1\%$ larger $D$). We can see (Fig.~\ref{fig7}B) that the energy density profile is indeed linear, as one would expect for a diffusive system with constant $D$. Likewise, the NESS current density is independent of site $p$ within about $1\%$ precision. Using system size $L=100$ and dividing the NESS energy current density value $j_p \approx 1.69\cdot 10^{-6}$ by the fitted gradient $\nabla h \approx 1.15\cdot 10^{-6}$, we get $D\approx 1.47$. Note that close to chain edges there are boundary effects on close to $20$ sites, which is perhaps related to the truncation support $r$ on which $\lambda_1$ converges (Fig.~\ref{fig7}A).

The second method studies the equilibrium energy current, specifically the autocorrelation function of the extensive current $J=\sum_p j_p$. According to the Green-Kubo formula~\cite{Pottier} the diffusion constant $D$ is
\begin{equation}
  D=\frac{1}{\chi}\int_0^\infty \lim_{L \to \infty}\frac{1}{L}\ave{J(t)J(0)}\,{\rm d}t,
\end{equation}
where $\chi=\tr{H_0' H_0'}/L={\cal J}^2+\hx^2+\hz^2$. Calculating correlation function $C_J(t)$ numerically exactly (Appendix~\ref{app:Num}) in a finite system of size $L=33$ we get $D \approx 1.45\pm 0.01$ (the error here stems from the fact that at finite $L$ such $C_J(t)$ agrees with the one in the TDL only upto a finite time $\approx 25$, see Fig.~\ref{fig7}C and Appendix~\ref{app:Num}) .

We therefore see that all three methods give the same value of $D$ within their $\approx 1-2\%$ accuracy, which furthermore agrees with the value reported in Ref.~\cite{white24}. We note that the chosen parameters studied in this part seem relatively ``easy'' in a sense that not too large $t$ or $L$ are necessary to get good convergence and accuracy. For other parameters, where for instance $C_J(t)$ oscillates and also takes negative values, things might be more difficult~\cite{foot2}.

\New{While our truncated propagator method scales exponentially with $r$ it still gives more precise values of $D$ with less numerical effort than other methods. Let us compare concrete numbers. Despite the gap being very small due to small $\tau=0.05$ and $k=0.02$, computation to get $\Delta$ takes $\approx 1.7\,$h at $r=10$, while at $r=13$ it takes $\approx 3$ days and requires $\approx 30$ Gb of memory. To get the NESS shown in Fig.~\ref{fig7}B with $L=100$ it on the other hand takes $\approx 19$ days, while to get the $L=33$ data for the current autocorrelation function in Fig.~\ref{fig7}C it takes $\approx 6$ days. Therefore, the truncated propagator method is significantly faster than other methods even using the largest $r=13$ where it results in much more precise $D$. At smaller $r=10-11$, where its precision would be comparable to the NESS method, it is $\sim 100$ times faster. Moreover, it seems that at least in the studied case, obtaining higher precision results is easier with the truncated propagator. To increase precision in Fig.~\ref{fig7}C, i.e., increase $L$ substantially beyond $33$, is nearly impossible due to prohibitive memory requirements (Appendix~\ref{app:Num}). To increase the precision of the NESS results one would likewise have to increase $L$. Due to finite-size corrections scaling as $1/L$~\cite{nesskubo} one would need $L \approx 400$ to decrease the error down to $0.005$ estimated for our truncated propagator method at $r=13$. This would be very time consuming becase in order to preserve precision of the MPO description we would also need a much larger bond size than $\chi=200$ (Appendix~\ref{app:Num}) rendering each relaxation step to reach NESS much slower. The reason that for a similar precision one on one hand needs $L\approx 400$ with NESS while already $r=13$ suffices for the truncated propagator method is because in the truncated operator method there are no finite-size effects at any $r$ -- it always works exactly in the thermodynamic limit -- and, furthermore, the convergence with $r$ seems to be rather fast (Fig.~\ref{fig7}A)~\cite{footconv}.}

\subsection{Prethermalization at large $\tau$ and DTC}
\label{sec:DTC}
\begin{figure}[t!]
  \centerline{\includegraphics[width=0.5\textwidth]{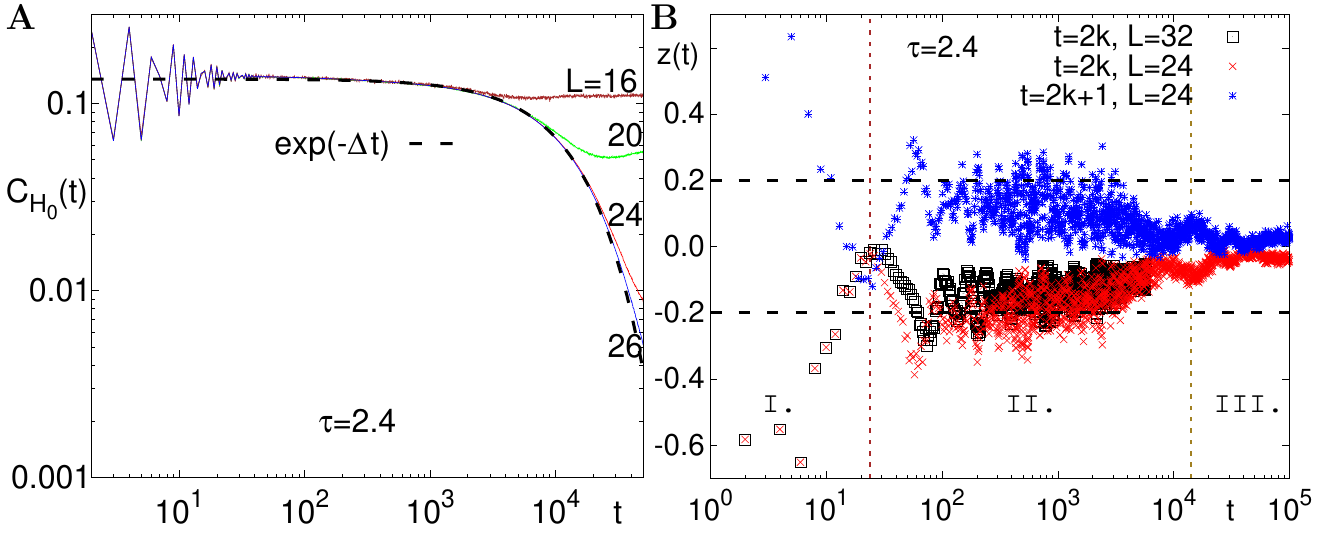}}
  \caption{{\bf Small frequency prethermalization and an almost DTC.} Prethermalization and discrete time crystal at large $\tau=2.40$. (A) correlation function at infinite temperature shows prethermalization. (B) Magnetization $z(t)=\ave{Z(t)}/L$, starting from a state $\ket{11\ldots 1}$ with $L=24$, plotted with different colors at odd and even steps, and $L=32$ (for clarity only even times are shown). First vertical brown line is at $t=24$ and separates regimes (see text) of fast decay and that of a prethermal DTC-like behavior, while the 2nd vertical line is at $1/\Delta$, indicating the end of prethermalization. \New{Data is for the kicked Ising model (\ref{eq:KIU}) with $J=\frac{1}{2}$, $\hx=\hz=1$.}}
  \label{fig8}
\end{figure}
In Fig.~\ref{fig2}C we can see that the gap is very small also for nonsmall $\tau$ (brown dashed curve) when one approaches the integrable point $\tau=\pi$. Let us explore if this is also an instance of prethermalization in the sense that there is an almost conserved operator and that appropriate correlation functions exhibit long prethermal plateau.

At the integrable point itself we can see that Eq.(\ref{eq:KIU}) simplifies to $U_{\rm x}=\prod_j (-\ii \sx_j)$, $U_{\rm z}=\prod_j (-\ii \sz_j)\prod_j(1-\ii \sz_j \sz_{j+1})/\sqrt{2}$, or all together
\begin{equation}
  U_\pi=(-1)^L \e^{-\ii\frac{\pi}{4}\sum_j \sz_j \sz_{j+1}}\e^{-\ii\frac{\pi}{2}\sum_j \sy_j }.
  \label{eq:Upi}
\end{equation}
This is nothing but the integrable kicked transversal field Ising model, e.g. Ref.~\cite{prosenPTPS}. In the presence of a disorder it would be an example of a much studied discrete time crystal (also the so-called $\pi$ spin glass)~\cite{vedika16,curt16,else16}. Note that our generator (\ref{eq:KIU}) with $\tau$ close to $\pi$ is not exactly equal to the one obtained by perturbing the angle $\pi$ in the two terms in Eq.(\ref{eq:Upi}), an often studied model of DTC~\cite{jaksch19,pizzi20,se21,marin21,deLuca}, however, as we shall see, its properties are rather similar. We also in particular have a specific value of the field in the $\sy$ direction that simplifies it even more (as well as a particular prefactor in front of the zz term~\cite{se21}). Namely, because the zz and the y terms commute, one can rewrite it as
\begin{equation}
  U_\pi^{2p}=\e^{-\ii \pi \frac{p}{2}\sum_j \sz_j\sz_{j+1}}.
\end{equation}
For odd $p$ each 2-site terms gives just a relative sign between the even and odd parity sectors, while for even $p$ it is identity, for instance, $U_\pi^4=\1$. If one stroboscopically looks at every $4$-th step the situation at $\tau=\pi$ is actually similar to the one at $\tau=0$: at $\tau=0$ one has trivial $U_0=\1$, whereas at $\tau=\pi$ the propagator $U_\pi$ is a 4-th root of identity.

Numerically calculating the correlation function, for instance of $H_0$ (\ref{eq:H0}, we indeed see (Fig.~\ref{fig8}) long prethermalization plateau, exactly as was the case for small $\tau$. The gap again perfectly predicts the thermalization time $T=1/\Delta$, with the scaling being (Fig.~\ref{fig2}C) $T \approx 10^{-8}\exp{(21/|\pi-\tau|)}$. This case of large $\tau$ is also interesting for another reason, namely, because the propagator is close to a root-of-identity we can expect an almost periodic behavior in time, that is, breaking of the time-translation symmetry, i.e., a discrete time crystal. It has been known~\cite{Nayak} that prethermalization is a way to obtain DTC in homogeneous systems, however not in (short-ranged) 1-dimensional systems. In our case we can observe that the total magnetization oscillates (Fig.~\ref{fig8}) with a period $2$. Similar behavior has been observed before in the kicked Ising model, see Refs.~\cite{marin21,jaksch19,pizzi20,se21}, seemingly at odds with a 1-dimensional nature of the system (for classical 1-d DTC see Ref.~\cite{li25}). We in fact have three different regimes in the behavior of $Z(t)$ (see Fig.~\ref{fig8}): (i) initial fast decay of magnetization until a minimum at $t \approx L$ (see Ref.~\cite{deLuca} for a detailed discussion of this regime), (ii) intermediate prethermal times $L \ll t \ll T$ where one has a DTC-like behavior, though the size of the plateau in the average spin magnetization $Z(t)/L$ here decays with $L$ as $\sim 1/L$, and (iii) asymptotic times $t \gg T$ where prethermalization breaks down and $Z(t)$ decays to a value that is exponentially small in $L$. Although one does not have a true DTC in the thermodynamic limit in regime (ii), it could still be of (experimental) interest as the plateau for specific initial states, like $\ket{1\ldots1}$, is still relatively large, say $>0.1$ at $L=32$, despite being rather far away from the exact DTC at $\tau=\pi$. Previous works~\cite{marin21,jaksch19,pizzi20,se21} mostly focused on smaller deviations from $\pi$, where one has to be extra careful about finite-size effects, see also discussion in Ref.~\cite{deLuca}. Because the truncated propagator gives an easy way to obtain the prethermalization duration $T=1/\Delta$ it also gives an exact prediction on the robustness of potential DTCs, avoiding confusing a finite-$L$ DTC with an exact one in the TDL. As discussed (Sec.~\ref{sec:Heis}), if one is interested in the TDL one should be aware of $T$ and of the Heisenberg time, and look at appropriate $L$ and $t$.

\begin{figure*}[t!]
    \centerline{\includegraphics[width=\textwidth]{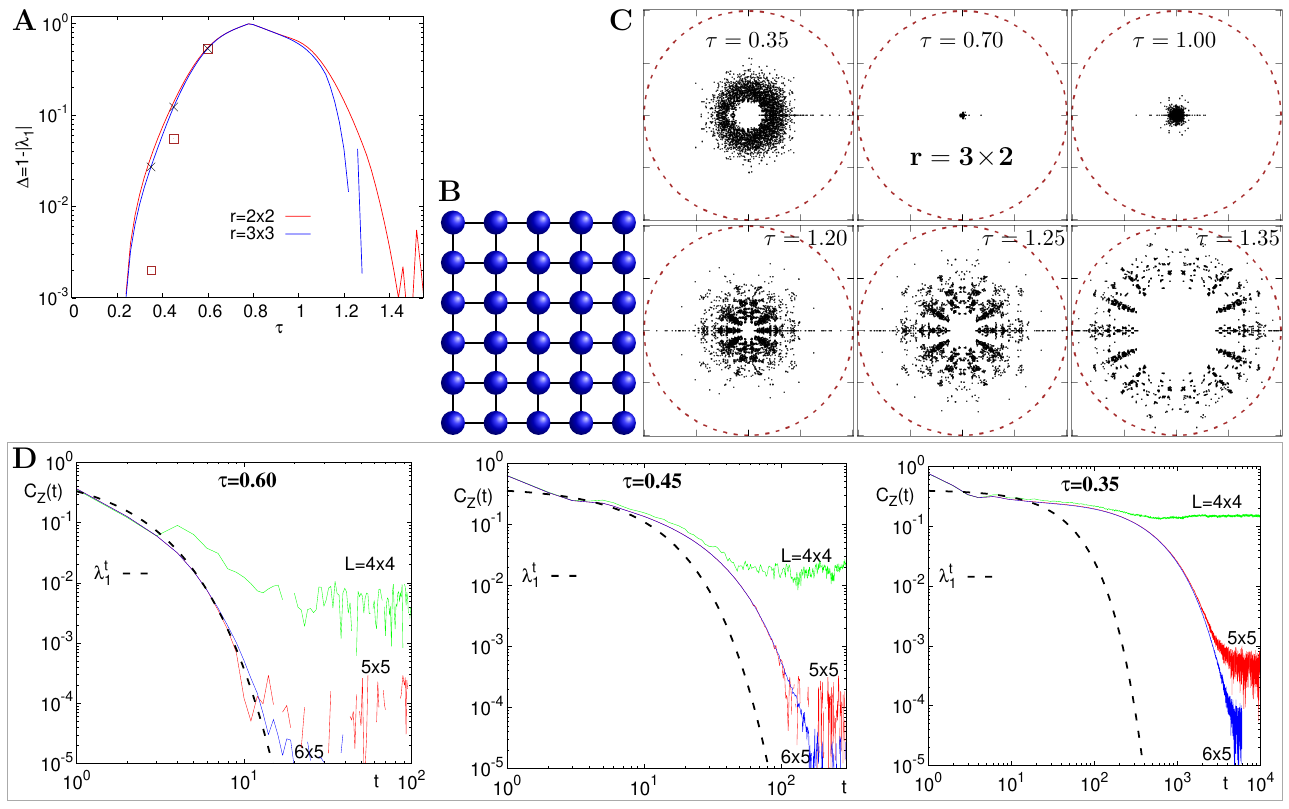}}
  \caption{\new{2D kicked Ising model on a square lattice (illustrated in (B)). (A) dependence of the gap on $\tau$. Three crosses mark the empirical decay used as dashed black exponential curves for three $\tau$ in (D), while brown squares are best fitted exponentials to numerical data for largest size $L=6\times5$ in (D). (C) shows spectra of $M(\mathbf{k}=0)$ for truncation $r=3\times2$.}}
  \label{fig:2d}
\end{figure*}
Worth noting is also that the scaling of the 2nd largest eigenvalues in different sectors (data like the one in Fig.~\ref{fig5}A) this time goes as $\sim |\pi-\tau|^3$, i.e., the power is $3$ rather than $2$. Finally, let us also note in passing that for the integrable kicked Ising model at $\tau=\pi$ (\ref{eq:Upi}) a local Floquet Hamiltonian (\ref{eq:HF}) does not exist despite the system not being ergodic, in fact, it is integrable with the two factors in Eq.(\ref{eq:Upi}) commuting, see Appendix~\ref{app:BCH} for details. While a standard argument for the absence of a local $\HF$ in chaotic Floquet systems is that they heat up~\cite{rigol14,lazarides14}, the reverse is obviously not true -- the absence of heating to infinite temperature does not imply the existence of a local $\HF$ (for a single-particle example see Ref.~\cite{haga19}).

\new{
  
\subsection{2D kicked Ising model}
So far we have demonstrated the ideas on the 1D kicked Ising model. The methods and conclusions though are model independent and here we briefly comment and show data for some other models.

One could specifically wonder if the kicked Ising model is generic enough, for instance, it has a single layer of commuting 2-qubit gates (Fig.~\ref{fig3}), whereas one might be interested in say a brickwall circuit where one has 2 layers of 2-qubit gates (with gates commuting within each layer). To that end we have implemented a model where on top of kicked Ising gates we add another layer of commuting $xx$ gates, thereby making the model more generic. In the limit of small $\tau$ such a model corresponds to a 1D XX model in a tilted magnetic field. More detailed results are presented in Appendix~\ref{app:C}, here we just summarize main findings. At small $\tau$ one again finds prethermalization with its timescale given by the leading Ruelle-Pollicott resonance of the truncated propagator. Close to $\tau=\pi$ things though are different than in the kicked Ising model. While the timescale again diverges, it does so only quadratically in $1/(\tau-\pi)$ (see Fig.~\ref{fig:xx}D), and not exponentially like in the kicked Ising model. This is so because the kicked XX model is not a DTC at $\tau=\pi$. In fact, behavior is similar to the kicked Ising model at small $\tau$ in sectors with nonzero quasimomentum $k$ where there are no almost-conserved operators (Fig.~\ref{fig5}). From the momentum dependence of $\lambda_1$ one can again get the energy diffusion constant which agrees with the one obtained from the integral of the current autocorrelation function via the Green-Kubo formula (Fig.~\ref{fig:xxdif}).

The truncated operator propagator can be used also in 2D models and we focus here on an example of a 2D kicked Ising model. Spins are now located on a 2D square lattice, with a one-step propagator being
\begin{eqnarray}
  U_\tau&=& U'_{\rm zz}  U'_{\rm z} U'_{\rm x},\\
  U'_{\rm x}&=&\prod_{i,j} \e^{-\ii \tau \hx \sx_{i,j}},\nonumber \\
  U'_{\rm z}&=& \prod_{i,j} \e^{-\ii \tau \hz \sz_{i,j}},\nonumber \\
  U'_{\rm zz}&=& \e^{-\ii \tau \sum_{\rm n.n.} \sz_{i,j} \sz_{i',j'}}, \nonumber
\end{eqnarray}
where the sum in the $U'_{\rm zz}$ is over nearest-neighbors on a 2D square lattice (Fig.~\ref{fig:2d}B). Similar to the 1D model we calculate the operator propagator in an infinite system truncated to a basis of local operators supported on $r$ sites. Because we are in 2D, the local basis operators are now from $r$ sites on an $n\times m=r$ grid. We show results for the largest eigenvalue for the $r=2\times2$ and $3 \times 3$ truncations in Fig.~\ref{fig:2d}(A). Because one has translational invariance in $x$ as well as in $y$ direction, resulting in two conserved quasimomentum components $k_x$ and $k_y$, construction of independent basis elements is a bit more involved. One has to ensure that all basis elements are independent, i.e., that they can not be obtained from each other by translations. In 1D it was enough to fix e.g. the first-site operator to be a non-identity, in 2D this is not enough. Details can be found in Appendix~\ref{app:2d}, let us just mention that the size of the basis approaches $4^{r=mn}$ for large $m$ and $n$. We also use in-situ numerical propagation of local basis operators so that we can easily study $r=3\times3$ truncation.

Dynamics of the 2D model probably warrants a separate study, here we focus on the main things in the zero quasimomentum sector ($k_x=k_y=0$), setting also $\hz=\hz=1$ and varying only $\tau$ (note that compared to the 1D kicked Ising model we do not use factors $1/2$ or $1/4$ in the exponents). At small $\tau$ one finds prethermalization with the prethermalization time being exponentially large in $1/\tau$, finding no Trotterization transition (Fig.~\ref{fig:2d}A). The decay time-scale at larger $\tau=0.60$ obtained from the exact numerical simulation of finite systems of size up-to $L=6\times5$ spins with periodic boundary conditions agrees with the one obtained from the truncated propagator at truncation size $r=3\times 3$ (Figs.~\ref{fig:2d}A,D). At smaller $\tau$ one can see a rapid increase of the prethermalization time, e.g. between $\tau=0.45$ and $\tau=0.35$ it increases by almost $2$ decades, though small linear truncation size ($2$ or $3$) is not yet large enough to get a precise value of the prethermalization time (this is actually similar as in 1D). Nevertheless, one can see a clear rapid decrease of the gap with decreasing $\tau$ (Fig.~\ref{fig:2d}A). Similarly as in 1D relying on small systems, e.g. with $L=4\times4=16$ spins, would incorrectly indicate a plateau at large times (green curves in Fig.~\ref{fig:2d}D).

\begin{figure}[t!]
    \centerline{\includegraphics[width=2.9in]{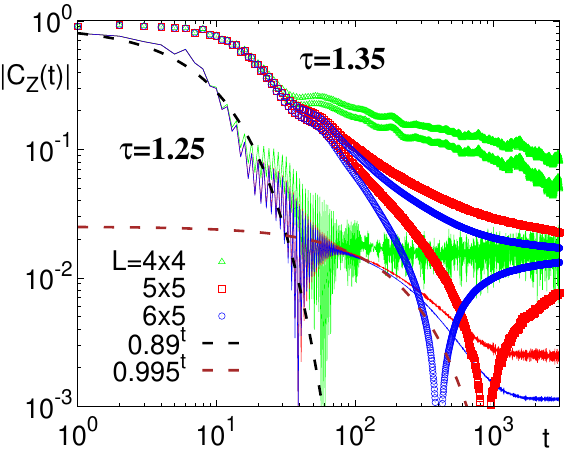}}
  \caption{\new{Transition in the behavior of magnetization correlation function in the 2D kicked Ising model around $\tau\approx 1.3$. Top set of data (points) is for $\tau=1.35$, while the lower ones (curves) are for $\tau=1.25$.}}
  \label{fig:2dC}
\end{figure}
There are two other interesting points worth highlighting. The first is that at $\tau=\pi/4$ all eigenvalues of $M(\mathbf{k}=0)$ are zero, i.e., the gap shown in Fig.~\ref{fig:2d}A is exactly one. While this is not specific to 2D, and has been already noted in the 1D kicked Ising model~\cite{RP24}, it means that all autocorrelation functions of extensive local operators decay immediately. We state some further properties of $M$ at $\tau=\pi/4$ and $\tau=\pi/2$ in Appendix~\ref{app:2d}.

The second point is an apparent transition in the dynamics around $\tau\approx 1.3$. Namely, for $\tau>\pi/4$ one notices that the largest absolute eigenvalue becomes negative, as opposed to being positive for $\tau<\pi/4$ (see e.g. the spectrum for $\tau=1.0$ in Fig.~\ref{fig:2d}C). As a consequence correlation functions will oscillate between negative and positive values between even and odd times. What is even more interesting is that around $\tau \approx 1.3$ there seems to be a qualitative change in the spectrum (Fig.~\ref{fig:2d}C) which does not anymore have a noisy ring or disk core (like in both 1D models). This change is reflected also in the decay of correlation functions. In Fig.~\ref{fig:2dC} we can see that at $\tau=1.25$ the autocorrelation function of magnetization has two distinct decay regimes: up-to $t \approx 50$ it decays fast in an oscillatory manner ($C_{\rm Z}$ at even and odd times have opposite signs), whereas for $t>50$ it is positive and decays slower -- perhaps still exponentially, though our largest system size is not large enough to verify that with confidence. On the other hand, for $\tau=1.35$ the decay is much slower, has a single regime ($C_{\rm Z}$ is always positive), with the asymptotic decay perhaps being more likely a power law rather than exponential. This could be due to a continuum of eigenvalues approaching $1$ in modulus (see Ref.~\cite{RP24} for an example where a continuum of eigenvalues (a branch cut) causes a power law decay).  

}

\section{Conclusion}

We have shown that the truncated propagator of extensive local operators is a very versatile method to study Floquet systems. It determines evolution of correlation functions, specifically, through the largest eigenvalue and the associated eigenvector it gives an explicit prediction for the prethermalization time and the effective (local) Floquet Hamiltonian governing prethermalization. This allowed us to resolve the question about reliability of approximating Hamiltonian evolution by a small-step Trotterization. Except for a single almost conserved observable all other orthogonal ones do not exhibit prethermalization, and even the almost conserved one decays in the thermodynamic limit. This means that due to the many-body dynamics the so-called shadowing time is always finite, and that there is no Trotterization transition in chaotic quantum systems. Previous observations of Trotterization transition were an artifact of finite-size effects. One has to ensure that the Heisenberg time is larger than the (long) prethermalization time. \new{Because the limits $L \to \infty$ and $t \to \infty$ do not commute, in order to get the correct thermodynamic results it is essential that our method works directly in the thermodynamic limit. If one would instead use small systems, for instance, exact diagonalization, one would get incorrect results.}

The truncated propagator can also be used to identify prethermal discrete time crystals. Just looking at the spectrum will already give an indication if there is a prethermal DTC (eigenvalues very close to roots of $1$, e.g., those around $-1$ in Fig.~\ref{fig2}C at $\tau=3.0$). Studying chaotic kicked Ising model we find, perhaps surprisingly, that the prethermalized regime takes up the larger part of the gate duration interval $\tau \in [0,\pi]$. For instance, prethermalization is absent (say gap $\Delta > 10^{-3}$, i.e. $T<10^3$) only for roughly $\tau \in [1.5,2.2]$. Glimpses of this rather slow thermalization in the 1D kicked Ising model can be seen already in data presented in Ref.~\cite{prosen07},\new{ and are present also in the 2D kicked Ising model}. Prethermalization seems to be more common than previously thought and is not limited only to large frequencies (small $\tau$).

Momentum dependence of the largest eigenvalue also contains information about transport. We have used it to calculate the energy diffusion constant during the \New{long} prethermalization phase, as well as for the Hamiltonian model obtained in the limit $\tau \to 0$. Despite simply using small $\tau$ to approximate Hamiltonian dynamics, which causes the gap to be very small and difficult to calculate, the obtained diffusion constant is very precise \New{and the method is faster than previously used methods}. 

While we always used numerics to obtain the spectrum of the truncated propagator, it would be interesting to see if exact solutions are possible for specific models. Coming up with a better numerical method that would allow one to get the relevant eigenvalues for truncation support cutoff larger than \new{about $r=14$ sites, feasible with the in-situ method that we use,} would be also a valuable extension. Note though that whether one has a prethermalization or DTC can be already inferred at small $r$, and so the method can be used as a simple tool to look for those regimes. In the kicked Ising model we in particular find a prethermalized DTC behavior for a wide range of parameters around $\tau=\pi$ (roughly $|\pi-\tau|< 0.9$), with the effect though slowly decaying with increasing system size.

\new{We have also studied the kicked XX model, where prethermalization is found only for small $\tau$, as well as 2D kicked Ising model finding an interesting transition in the decay of correlation functions.}

\section*{Acknowledgments}
The author would like to thank Urban Duh for work on related project~\cite{urban}, and acknowledge Grants No.~J1-4385, No.~J1-70049 and No.~P1-0402 from Slovenian Research Agency (ARIS).


\newpage

\appendix

\section{Numerics}
\label{app:Num}

\subsection{Spectrum of $M(k)$}

The size of matrix $M(k)$ is $N=\frac{3}{4} 4^r$. For smaller $r$ the size is small enough to numerically diagonalize the whole matrix. For larger $r$ one can use methods targeting only the largest eigenvalue (in modulus), or few largest, like the simple power method for $\lambda_1$, or Lanczos based methods (we have used the ARPACK package~\cite{arpack}) that construct a tridiagonal matrix representing $M$ on a Krylov subspace. The method does not require the whole matrix, which would be memorywise prohibitive, but rather needs only application of $M$ on a vector. This can be simply done by simulating the action of our circuit on $r+2$ sites, see Eq.(\ref{eq:Mk}), that is, consecutively applying individual gates (Fig.~\ref{fig3}). \new{While the implementation on $r+2$ sites is straightforward, one can improve on speed and memory by directly performing evolution in-situ on just $r$ sites as described in the next subsection.}

\begin{figure}[h!]
  \centerline{\includegraphics[width=0.35\textwidth]{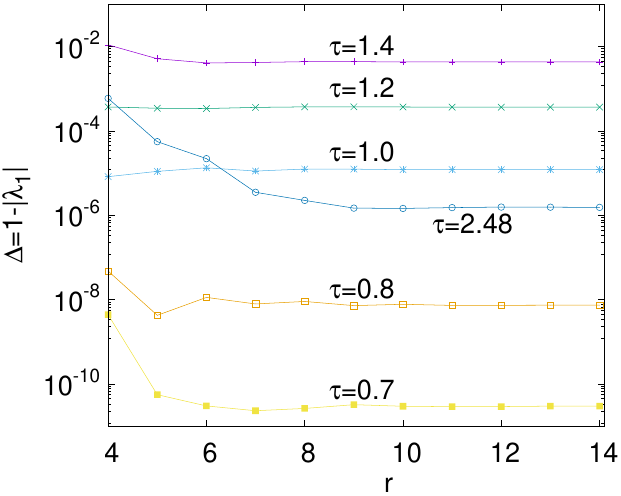}}
  \caption{Convergence of the gap with $r$ for selected $\tau$ (symbols in Fig.~\ref{fig2}C).}
  \label{fig9}
\end{figure}

In Fig.~\ref{fig9} we check convergence of $\lambda_1(k=0^+)$ that determines the prethermalization time with the support size $r$, confirming that $T$ reported in Fig.~\ref{fig2}C is the asymptotic one.

\new{
\subsection{In-situ propagation of operators}
\label{app:insitu}
To get the action of the truncated propagator we need the overlaps of the propagated local operator $U^\dagger \a{{\beta}}_1 U$, where $\a{{\beta}}_1$ lives on $r$ sites, with local operators that are either on the same $r$ sites, or are shifted by one site to the left or to the right (for evolutions that can shift by at most $1$ site, like the kicked Ising model), see Eq.~(\ref{eq:Mk}). A simple way to do that is simply to evolve the operator $\a{{\beta}}_1$ on $r+2$ sites, starting by padding one identity at the left and the right edge, calculating the full $U^\dagger \a{{\beta}}_1 U$ that now lives on $r+2$ sites, and then projecting it to the relevant operators with support on $r$ sites. However, that is expensive: we have to do evolution on $r+2$ sites and each additional site increases memory and computational requirements by a factor of $\approx 4$. Much faster is to do the in-situ evolution on just the relevant $r$ sites. In the following we describe how to do that, having in mind the 1d kicked Ising model.

One uses the fact that an initial operator with support on $r$ sites can increase its support only at its edges. Specifically, to get the contribution without any shift, $\a{{\beta}}_1 \to \a{{\alpha}}_1$, the two qubit gate acting on the boundary (dashed lines in Fig.~\ref{fig:skicar}(a)) has to map at the left edge $\mathbbm{1}_0 x_1$, where ``$x$'' indicates any non-identity operator, $x \in \{ \sx,\sy,\sz\}$, to again the identity at the leftmost site (see Fig.~\ref{fig:skicar}(a)). For the kicked Ising model where the 2-qubit gate has only the $zz$ interaction the only possibility is $\mathbbm{1}_0 x_1 \to \mathbbm{1}_0 x_1$. At the right edge one has a similar situation, with the only difference that the last operator can be also the identity, resulting in $(*)_r \mathbbm{1}_{r+1} \to (*)_r \mathbbm{1}_{r+1}$, where $(*) \in \{ \sx,\sy,\sz,\mathbbm{1}\}$.
\begin{figure}[h!]
    \centerline{\includegraphics[width=3.2in]{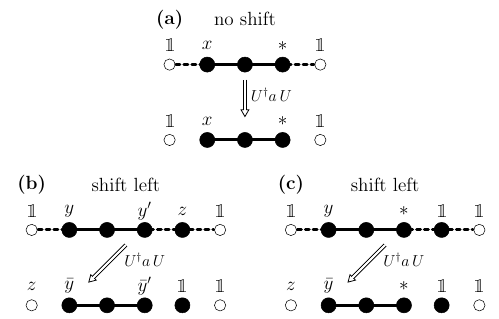}}
  \caption{\new{Truncated propagator construction: in-situ propagation of local operators on just $r$ sites. The initial operator is supported on $r$ sites (black circles) padded with an identity outside its support (white circles). Executing a 2-qubit $zz$ gate on dashed lines has to execute a mapping indicated by the arrow. Afterwards one has to perform the remaining 2-qubit gates (full lines). See text for details.}}
  \label{fig:skicar}
\end{figure}
If we want to achieve a mapping of the initial operator to a shifted one supported on $r$ consecutive sites we have to get an identity at its tail end, and a non-identity operator at its front-end. For the $zz$ gate in question such a mapping can be achieved only by $\mathbbm{1}_0y_1 \to \sz_0\bar{y}_1$, where ``$y$'' stands for $y \in \{ \sx,\sy\}$, and its negation $\bar{y}$ is $\sx$ if $y=\sy$ and $\sy$ if $y=\sx$. This is sketched in Fig.~\ref{fig:skicar}(b). For a shift to the left there is another possibility, namely, the right-most site can already have an identity operator, resulting in a scheme in Fig.~\ref{fig:skicar}(c) (for a shift to the right (b) is the only option). We can see that in both Fig.~\ref{fig:skicar}(b-c) it is enough to keep track of only $r$ sites -- at the beginning on $r$ black circles, and after performing the first step (transformation on all dashed links) only on the leftmost $r$ sites, afterwards executing the remaining gates (full-line links in Fig.~\ref{fig:skicar}).
}

\subsection{Correlation functions}

We calculate correlation functions in finite systems numerically exactly by evolving a pure random initial state $\ket{\psi}$ in up-to $L=33$ spins by repeatedly applying one and 2-qubit gates (see Fig.~\ref{fig3}). For large $L$ such a single random state expectation value is a good approximation for an infinite temperature average. Namely, finite-size fluctuations in a correlation function scale as $\sim 1/2^{L/2}$, and are e.g. $\sim 10^{-5}$ for $L=32$. We always look at values of $C_A(t)$ that are larger than that (e.g. Fig.~\ref{fig5}B). Note that in some cases, like studies of transport, there can be other boundary finite-size effects that are larger, see Fig.~\ref{fig7}C for an example. Therefore we always check convergence with $L$, making sure to have finite-size effect under control. Memory requirements of the method are exponential in $L$ (keeping track of two pure states, each of size ${\cal N}=2^L$), and reach $\approx 400\, $Gb for $L=33$, while each gate application takes ${\cal O}({\cal N})$ operations.

\begin{figure}
    \centerline{\includegraphics[width=0.5\textwidth]{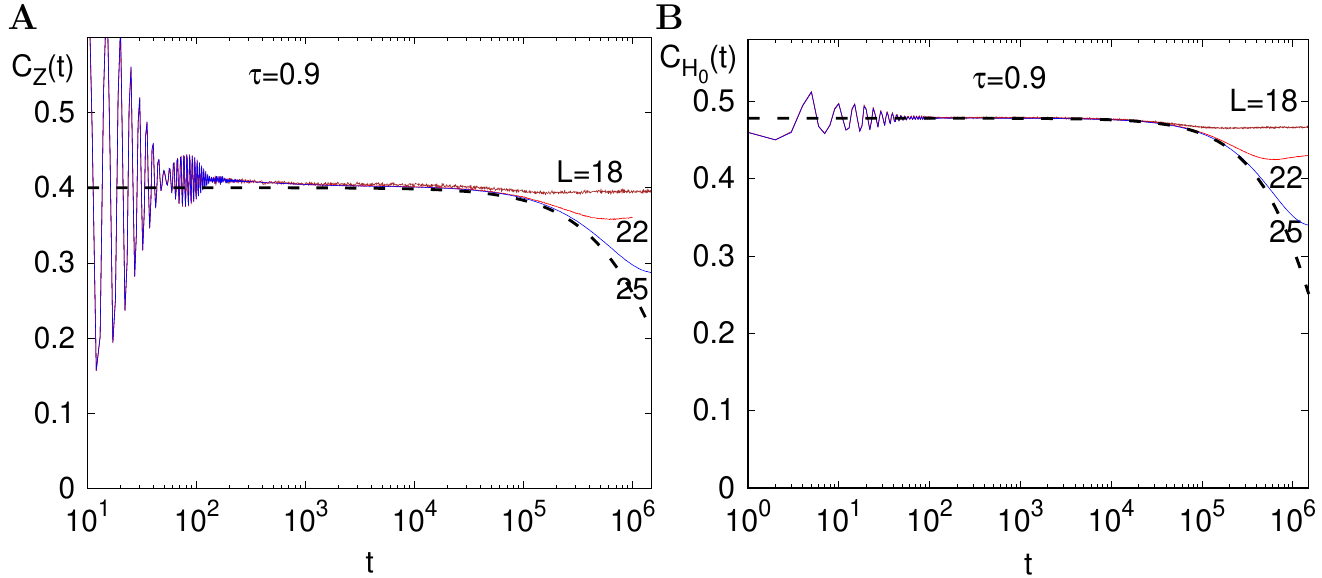}}
  \caption{Autocorrelation functions at $\tau=0.9$ again decay to zero in the TDL. Theoretical dashed lines (no fitting parameters) is exponential decay agreeing with numerics for large $L$. \New{All for kicked Ising model (\ref{eq:KIU}) with $J=\frac{1}{2}$, $\hx=\hz=1$.}}
  \label{fig10}
\end{figure}
\begin{figure*}
    \centerline{\includegraphics[width=0.98\textwidth]{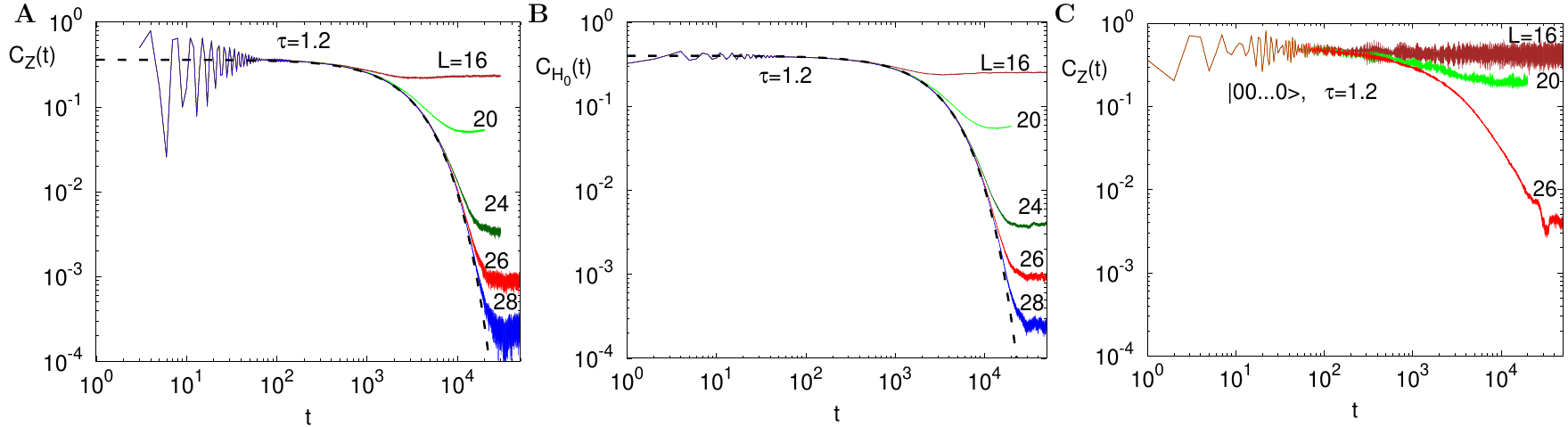}}
  \caption{Autocorrelation functions at $\tau=1.2$ and random initial states (A and B), as well as for a polarized initial state $\ket{00\cdots 0}$ in C. \New{All for kicked Ising model (\ref{eq:KIU}) with $J=\frac{1}{2}$, $\hx=\hz=1$.}}
  \label{fig11}
\end{figure*}

In Fig.~\ref{fig10} and Fig.~\ref{fig11} we check that correlation functions decay to zero in the TDL also for other values of $\tau$. At $\tau=0.9$ the prethermalization time is so large that the simulation takes a lot of time and we can not afford to simulate as large systems. Nevertheless, $L=25$ already reveals that the same prethermalization theory still applies, i.e., there is no Trotterization transition, despite being well below the claimed transition (crossover) point in Ref.~\cite{Heyl}. At $\tau=1.2$ in Fig.~\ref{fig11} the decay is rather clear, including for a specific fully polarized initial state.

In calculating the energy current autocorrelation function (\ref{eq:J}) in Fig.~\ref{fig7}C we have simply used the kicked Ising model with a small timestep $\tau=0.05$. Comparing results at $L=32$ between $\tau=0.05$ and $\tau=0.02$ we can see that e.g. up-to $t=30$ the difference in $D$ is about $0.002$ due to a finite Trotterization step approximation of Hamiltonian evolution. On the other hand, comparing different $L$ at $\tau=0.05$ we can estimate that the $L=33$ results do correctly reproduce the correlation function in the thermodynamic limit up-to $t \approx 25$ with precision in $D$ of about $0.002$ (this time scales diffusively $\propto L^2$ with system size). Because $C_J(t)$ at $t=25$ is already quite small, taking into account that the remaining integral from $t=25$ to $\infty$ would bring only about $\approx 0.01$, we can estimate the diffusion constant by the integral of $L=33$ data up-to $t=30$, which gives $1.452$, and estimate error at $0.01$.

One can in fact use the Green-Kubo integral up-to a finite time $t$ up-to which we trust to have the correct $C_J(t)$ in the TDL as a lower bound on $D$ because $C_J(t)$ for the chosen parameters is always positive, thereby estimating $D>1.45$. On the other hand, convergence of the truncated operator propagator method (Fig.~\ref{fig7}A) seems to be monotonously from above, so using the $r=12$ value of $D=1.462$ we can estimate $D<1.46$. Both methods together therefore suggest tighter estimate $D=1.45-1.46$.

\subsection{Boundary driven systems and NESS}

To obtain a NESS and study transport properties of the chaotic Ising model $H_0$, specifically its diffusion constant, we use a boundary driven Lindblad equation. The method has been used many times since its first use on the very same Ising model~\cite{jstat09} (for different parameter values). Let us briefly outline its working.

Evolution of the density operator $\rho(t)$ is described by the Lindblad master equation,
\begin{equation}
\frac{{\rm d}\rho}{{\rm d} t}=i [\rho,H]+\sum_j([L_j \rho,L_j^{\dagger}]+[L_j, \rho L_j^{\dagger}]),
\end{equation}
where $L_j$ are Lindblad operators. For our simulations we want to induce the flow of energy in the NESS and the Lindblad operators act on two boundary spins at each end. They are chosen in such a way that the dissipator ${\cal L}$ given solely by Lindblad operators on those two spins would induce a canonical state of two boundary spins. More precisely, $L_j$ at say the left edge are such that ${\cal L}(\rho_{L})=0$, where $\rho_L$ is the reduced density operator obtained by tracing over $4$ out of $6$ spins, $\rho_L \sim \tr{\e^{-H/T_L}}$, where $H$ is the Hamiltonian on $6$ spins. See Ref.~\cite{jstat09} for more details. We use Lindblad (bath-like) ``temperature'' parameters $T_L=295$ and $T_R=300$, inducing an almost infinite-temperature NESS in which the average energy density is small.

The NESS, which is unique, is obtained by evolving $\rho(t)$ in time (Trotter step in the 4th order scheme is $0.05$) until convergence to the NESS is obtained (for $L=100$ this relaxation time is $5\cdot 10^3$). Evolution is performed by writing $\rho$ in a matrix product operator (MPO) form with matrices of maximal bond size $\chi$ ($\chi=200$ for data in Fig.~\ref{fig7}B), \New{with numerical complexity scaling as ${\cal O}(\chi^3)$}. Application of a single two-site gate is performed using standard time-evolved-block decimation (TEBD) techniques~\cite{vidal04}. Convergence with the MPO bond dimension seems to be rather slow (definitely slower than in many other studies of spin transport with the same method). The value of diffusion constant that we get, $D \approx 1.47$, seems to be a bit on the larger side compared to the other two methods.

\section{BCH series}
\label{app:BCH}

Writing all propagators in $U=U_{\rm z}U_{\rm x}$ in terms of generators as
\begin{equation}
  U_{\rm z}={\rm e}^{-\ii A \tau},\quad  U_{\rm x}={\rm e}^{-\ii B \tau},\quad U={\rm e}^{-\ii H_{\rm F}\tau},
\end{equation}
the Baker-Campbell-Hausdorff (BCH) series gives
\begin{equation}
  H_{\rm F}=A+B+\frac{(-\ii \tau)}{2}E_3+\frac{(-\ii \tau)^2}{12}([A,E_3]-[B,E_3])+\cdots,
    \label{eq:bch12}
\end{equation}
where $E_{3}\equiv [A,B]$, and we explicitly wrote out terms up-to to 3rd order in products of $A$ and $B$. More systematically, the series can be written in such a way that all next order terms are a single commutator of two terms occurring in previous orders, i.e., 
\begin{equation}
  H_{\rm F}=\sum_{j \ge 1} (-\ii \tau)^{|j|-1} z_j E_j \equiv \sum_{m=0} (-\ii \tau)^m H_m,
  \label{eq:BCHorders}
\end{equation}
where $E_j\equiv [E_{j'},E_{j''}]$ with appropriate $j',j''<j$, starting with $E_1\equiv A, E_2\equiv B$ and $|1|=|2|=1$, while $z_j$ are rational numbers. Coefficients up-to order $20$ are provided in Ref.~\cite{casas08} ($|j|\le 20$, i.e., $m\le 19$ in Eq.\ref{eq:BCHorders}). For instance, the highest order $|j|=20$ contains $52377$ terms $E_j$, while all $20$ orders together have ${\rm max}(j)= 111013$ terms. 

\begin{figure}
    \centerline{\includegraphics[width=0.5\textwidth]{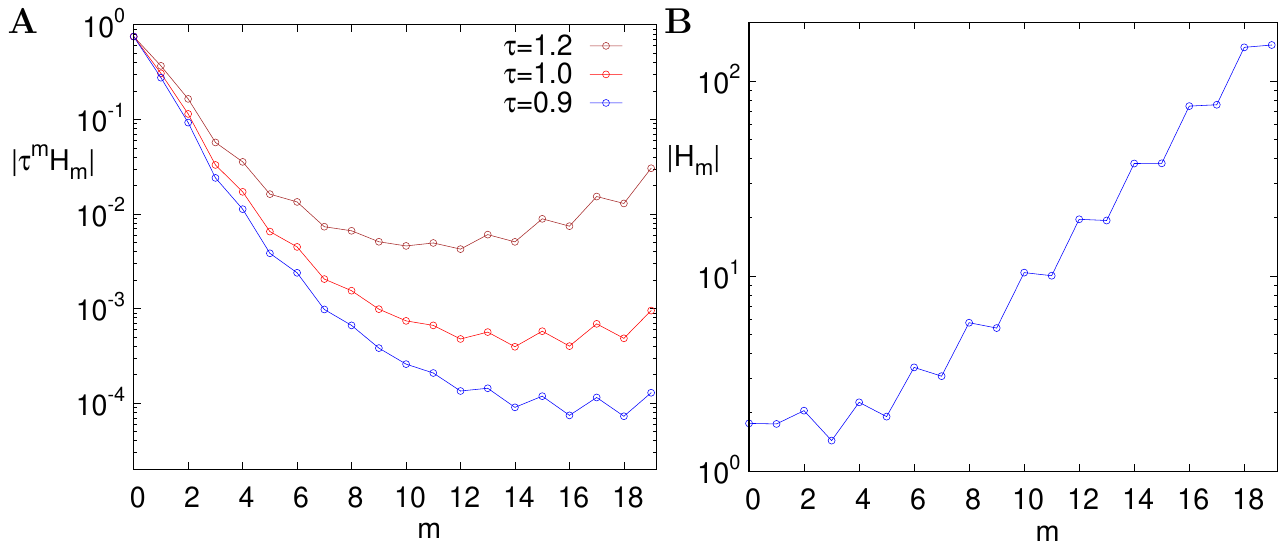}}
   \caption{{\bf Divergence of the BCH series for the kicked Ising model.} (a) Chaotic case in the prethermalization regime. (b) Integrable case at the DTC point $\tau=\pi$. In all cases (\ref{eq:KI}) we use $J=1/2, \hx=\hz=1$.}
   \label{fig12}
\end{figure}

We have symbolically calculated all $H_0,\ldots,H_{19}$ in an infinite system, each $H_m$ being a translational sum of densities, $H_m=\sum_{\boldsymbol \alpha}c^{(\boldsymbol{\alpha})}\sum_j \sigma_j^{(\boldsymbol{\alpha})}$, with the Hilbert-Schmidt (i.e., Frobenius) norm being $|H_m|^2=\sum_{\boldsymbol \alpha} (c^{(\boldsymbol{\alpha})})^2$. Norms of individual orders are shown in Fig.~\ref{fig12} for different $\tau$. Note that the $\tau$ dependence of $\HF$ is trivial (\ref{eq:BCHorders}); the non-trivial part are $H_m$. A similar plot has been shown for different parameters in Ref.~\cite{RP24}, and for a finite system in Ref.~\cite{Mori16}.

We can also calculate the prethermalization finite-time plateau. If $H_{\rm F}'$ would be conserved (which is not), the plateau in energy correlation function of $H_0$ would be equal to the square of the projection of $H_0$ to $H_{\rm F}'$, which can be written as
\begin{equation}
  \ave{C_{H_0}}=\frac{(\tr{H_0 H_{\rm F}})^2}{\tr{H_{\rm F}^2}}.
\end{equation}
Evaluating traces to few lowest orders in $\tau$ we get ($J=\frac{1}{2},\hx=\hz=1$)
\begin{equation}
  \ave{C_{H_0}}\approx \frac{9}{16}-\frac{3}{32}\tau^2-\frac{13}{1152}\tau^4+\cdots,
\end{equation}
for instance evaluating to $\ave{C_{H_0}}\approx 0.457$ for $\tau=1.0$ used in Fig.~\ref{fig1}D. The leading term above is simply $\tr{H_0^2}=J^2(\hx^2+\hz^2+\frac{1}{4})$, which would be the plateau in the autonomous limit.

Doing similar projection for the magnetization correlation function gets
\begin{equation}
  \ave{C_Z}=\frac{(\tr{Z H_{\rm F}})^2}{\tr{H_{\rm F}^2}}\approx \frac{4}{9}-\frac{4}{81}\tau^2-\frac{73}{14580}\tau^4+\cdots,
\end{equation}
evaluating to $\ave{C_Z}\approx 0.39$ for $\tau=1.0$ shown in Fig.~\ref{fig1}D. Again, the leading order term, obtained for the Hamiltonian limit $\tau=0$, is simply $\hz^2/(\hz^2+\hx^2+1/4)$.

\begin{figure*}[t!]
    \centerline{\includegraphics[width=\textwidth]{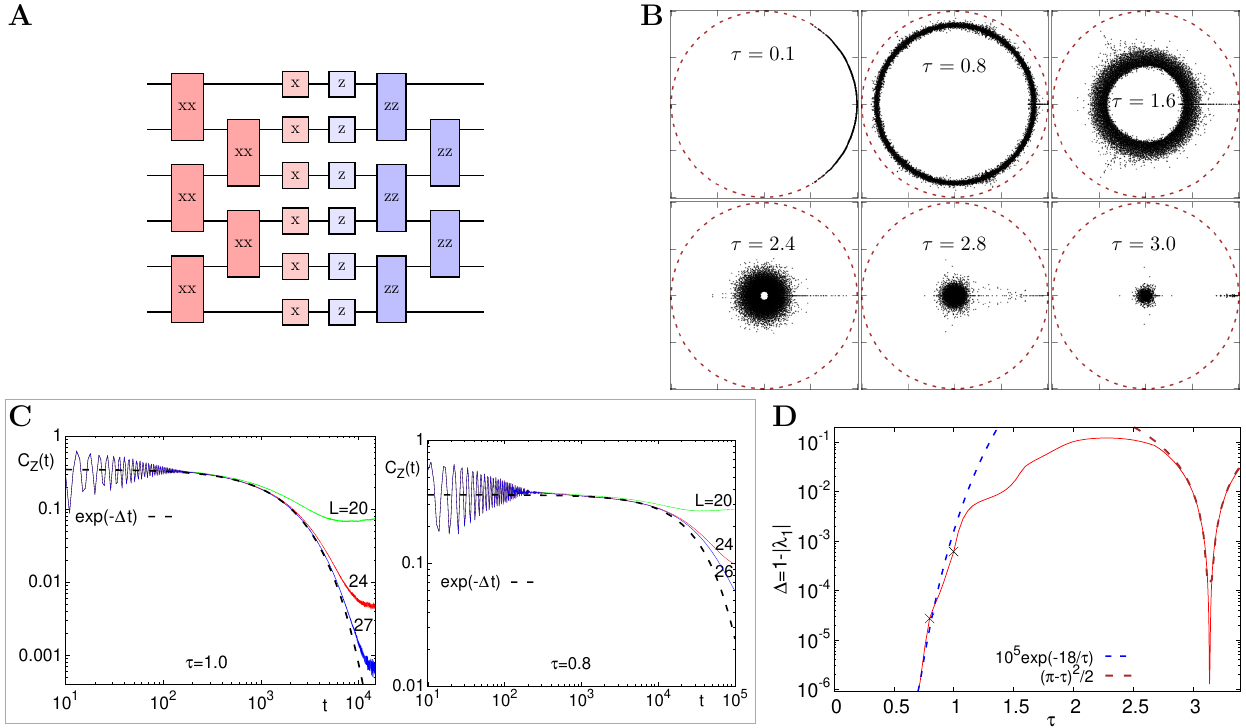}}
  \caption{\new{1D kicked XX model (\ref{eq:kxx}). (A) shows the circuit, (B) the spectra of truncated propagators for $r=7$, (C) autocorrelation function of magnetization (black dashed curves use the gap from (D)), and (D) the gap for $r=8$. Two crosses indicate the decay rate for dashed curves for $\tau$ in (C).}}
  \label{fig:xx}
\end{figure*}
\subsection{Integrable transversal Ising}

The integrable kicked Ising propagator at $\tau=\pi$ is very special; it can be written (\ref{eq:Upi}) as $U_\pi=(-1)^L \e^{\ii A}\e^{\ii B}$ ($\tau$ is here included in $A$ and $B$), with the two unitary factors commuting. Despite that, it can not be written as $U_\tau=\e^{-\ii \HF \tau}$ with a local Floquet Hamiltonian $\HF$. Namely, we have calculated successive orders in the BCH series in an infinite system for this special transversal Ising model, seeing a clear divergence with $m$ (Fig.~\ref{fig12}B). The structure of different orders is here simple. At even order $m=2k$ there are exactly $2(k+1)$ nonzero terms in $H_m$, namely: one $\sy$, $k+1$ terms of form $\sz(\sy\cdots\sy)\sz$ with a varying number of consecutive $\sy$ (from $0$ to $k$), and $k$ terms of form $\sx(\sy\cdots\sy)\sx$ with the number of consecutive $\sy$ varying from $0$ to $k-1$ (a structure inherited from Jordan-Wigner transformation). For odd $m=2k+1$ one has $2(k+1)$ nonzero Pauli strings, namely $\sx(\sy\cdots )\sz+\sz(\sy \cdots )\sx$, with the number of $\sy$ going from $0,1,\ldots,k$. $\HF$ is therefore nonlocal in Pauli operators. We also remind (Sec.\ref{sec:KI}) that at $\tau=\pi$, regardless of $k$, the truncated propagator has all nonzero eigenvalues equal to $|\lambda_j|=1$ (a fraction $5/24$ of eigenvalues is $+1$, and the same number $-1$), i.e., all eigenvectors are either exactly conserved by $U^2_\pi$, or are from the kernel (a fraction $7/12$ of all operator space size).

\new{
\section{1D kicked XX model}
\label{app:C}

The one-step propagator we take is (Fig.~\ref{fig:xx}A)
\begin{eqnarray}
    \label{eq:kxx}
  U_\tau&\equiv& U_{\rm zz} U_{\rm z} U_{\rm x} U_{\rm xx},\\
  U_{\rm x}&=&\prod_j \e^{-\ii \tau \hx \sx_j/2},\nonumber \\
  U_{\rm z}&=& \prod_j \e^{-\ii \tau \hz \sz_j/2},\nonumber \\
  U_{\rm zz}&=&\e^{-\ii \tau \sum_j \sz_j \sz_{j+1}/4}, \nonumber\\
  U_{\rm xx}&=&\e^{-\ii J_{\rm x}\tau\sum_j \sx_j \sx_{j+1}/4}. \nonumber
\end{eqnarray}
Compared to the kicked Ising model we have an additional layer of commuting $xx$ gates. We use $\hx=\hz=1, J_{\rm x}=1$. Results for the spectrum of the truncated propagator, correlation function in prethermalization regime of small $\tau$, and dependence of the gap on $\tau$ are shown in Fig.~\ref{fig:xx}. As discussed in the main text, one notable difference compared to the 1D kicked Ising model is behavior close to $\tau=\pi$. Here the timescale is only quadratically large in $1/|\tau-\pi|$, rather than exponentially as in the kicked Ising model. This is because at $\tau=\pi$ the kicked XX model is not a DTC, i.e., all nontrivial eigenvalues of $M$ go towards $\lambda=1$, as opposed to the DTC behavior in the 1D and 2D kicked Ising models where they go towards $\pm 1$.

\subsection{Energy diffusion}
We can again study energy transport in the Hamiltonian limit by taking small $\tau$, where dynamics approximates Hamiltonian
\begin{equation}
  H_0'=\sum_j {\cal J}_{\rm z} \sz_j\sz_{j+1}+{\cal J}_{\rm x}\sx_j \sx_{j+1}+\hz \sz_j+\hx \sx_j.
  \label{eq:H0c}
\end{equation}
In numerics we will use ${\cal J}_{\rm z}={\cal J}_{\rm x}=\hx=\hz=1$. The energy current is
\begin{eqnarray}
  j_{p}&=&\hx{\cal J}_{\rm z}( \sy_{p} \sz_{p+1}-\sz_{p-1} \sy_{p})+\\
  &+&\hz {\cal J}_{\rm x} (\sx_{p-1}\sy_{p}-\sy_{p}\sx_{p+1})+\\ \nonumber
  &+&2{\cal J}_{\rm z} {\cal J}_{\rm x} (\sx_{p-1}\sy_{p}\sz_{p+1}-\sz_{p-1}\sy_{p}\sx_{p+1}).
  \label{eq:Jxx}
\end{eqnarray}
Results for the diffusion constant $D$ obtained from the truncated propagator using $\tau=0.05$ and $k=0.01$ and $k=0.02$ are in Fig.~\ref{fig:xxdif}(a), and agree with $D$ obtained by evaluating the integral of the energy current autocorrelation function shown in Fig.~\ref{fig:xxdif}(b). The energy current autocorrelation function is at first sight similar to the one for the 1d kicked Ising model in Fig.~\ref{fig7}C, the difference here being more regular oscillations visible at smaller $t$.
\begin{figure}[h!]
    \centerline{\includegraphics[width=3.1in]{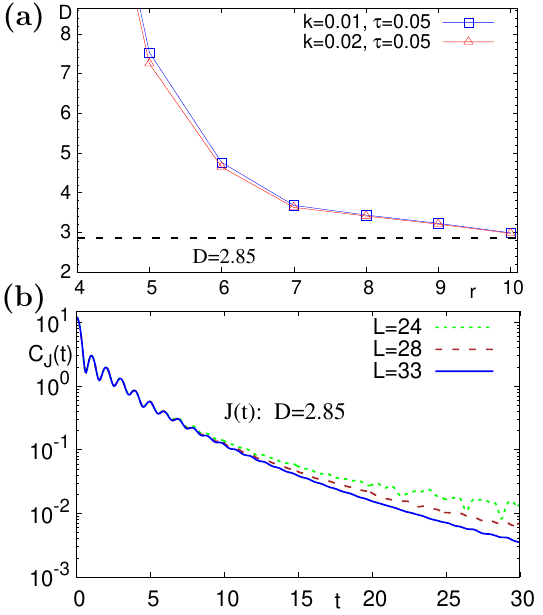}}
  \caption{\new{Energy diffusion constant in the XX model (\ref{eq:H0c}). (a) from the truncated propagator, (b) from current autocorrelation function.}}
  \label{fig:xxdif}
\end{figure}
}

\new{

  \section{2D kicked Ising model}
  \label{app:2d}
  
\subsection{Local operator basis}

Let us illustrate the local basis construction on a simple case of a $r=2\times2=4$ sites. All basis elements can be grouped into individual classes according to the position of all identity operators. For $r=2\times2$ there are $N_{\rm cl}=10$ classes that we list as symbols arranged in a $2\times 2$ pattern indicating their position in a 2D grid, $1$ denoting the identity operator and ``$x$'' an arbitrary non-identity Pauli matrix,
\begin{equation}
  \left(
    \begin{smallmatrix}
      1 & 1 \\
      1 & x
    \end{smallmatrix},
    \begin{smallmatrix}
      1 & 1 \\
      x & x
    \end{smallmatrix},
    \begin{smallmatrix}
      1 & x \\
      1 & x
    \end{smallmatrix},
    \begin{smallmatrix}
      x & 1 \\
      1 & x
    \end{smallmatrix},
    \begin{smallmatrix}
      1 & x \\
      x & 1
    \end{smallmatrix},
    \begin{smallmatrix}
      x & x \\
      1 & x
    \end{smallmatrix},
    \begin{smallmatrix}
      1 & x \\
      x & x
    \end{smallmatrix},
    \begin{smallmatrix}
      x & 1 \\
      x & x
    \end{smallmatrix},
    \begin{smallmatrix}
      x & x \\
      x & 1
    \end{smallmatrix},
    \begin{smallmatrix}
      x & x \\
      x & x
    \end{smallmatrix}
    \right)
    \label{eq:basis}
\end{equation}
The number of basis elements in a given class is equal to $3^p$, where $p$ is the number of ``$x$'' symbols, resulting in the total basis size for $2\times2$ truncation being $N=228=\sum_j^{N_{\rm cl}} 3^{p(j)}$. All other arrangements of operators on $2\times 2$ sites, for instance $\left(\begin{smallmatrix} x & 1\\x &1\end{smallmatrix}\right)$ that is not a member of the above basis, can be obtained from the basis in Eq.(\ref{eq:basis}) by translations in $x$ and $y$ directions. Compared to the 1D case basis we can see that we have $2$ additional classes (5th and 9th in Eq.(\ref{eq:basis})) in which the lower rightmost operator is not necessarily a non-identity.

For larger truncation sizes $r$ procedure is conceptually the same but combinatorially more complicated, however basis can be easily constructed using translations to eliminate elements. We just list the basis sizes and the number of classes in Table~\ref{tab:baza} .
\begingroup
\squeezetable
\begin{table}[t!]
  \begin{ruledtabular}
  \begin{tabular}{rccccc}
\multicolumn{1}{l}{} & \multicolumn{5}{c}{Truncation $r$} \\
\cmidrule(r){2-6} & $2\times2$ & $3\times2$ & $3\times3$ & $4\times 3$ & $4\times4$\\
\midrule
{\rm basis size} $N$ & 228   & 3792 & 254\,208 & 16\,453\,632 & 4\,261\,675\,008\\
{\rm classes} $N_{\rm cl}$     & 10 & 44 & 400 & 3392 & 57\,856\\
\end{tabular}
\end{ruledtabular}
  \caption{\new{The local operator basis sizes for different supports $r$ in 2D square lattice with translational invariance by one site, e.g., in 2D kicked Ising model.}}
         \label{tab:baza}
\end{table}
\endgroup
For an $r=n \times m$ truncation support the basis size is a bit less than $4^{nm}$, while the number of classes is upper bounded by $2^{nm}$ (for large $n$ and $m$ it approaches these two bounds, while the exact formula for the number of classes for e.g. $r=n\times 2$ basis is $(5+6(2^{n-2}-1))2^{n-1}$, and for $r=n\times 3$ it is $(11+14(2^{n-2}-1))2^{2n-2}$).

\begin{figure}[b!]
  \includegraphics[width=2cm]{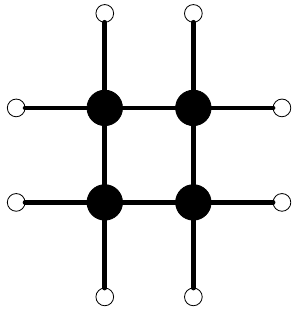}
  \caption{\new{For $r=2\times 2$ truncation and 2D kicked Ising model the initial operator supported on black circles can spread at most by 1 site in each direction (white circles).}}
  \label{fig18}
\end{figure}
To obtain $M$ one needs to propagate the local basis. An easy way is to, similarly as in 1D, include in propagation all sites directly connected to the $r$ basis sites (a layer of commuting 2-site gates, like in kicked Ising models, can increase support only by one site to its neighbors). For instance, for a $r=2\times 2$ basis this requires $8$ additional sites (Fig.~\ref{fig18}), demanding in total propagation on $r+8=12$ sites. While this is numerically feasible, larger bases would be not. Therefore we for 2D kicked Ising model use an in-situ propagation on just $r$ sites. This allows us to treat also $r=3\times 3$ basis with ease. The in-situ propagation for a 2D support is a bit more nuanced than in 1D, but follows the same idea.

\subsection{Special point $\tau=\pi/4$}

One has $\e^{-\ii \frac{\pi}{4}\sz_j \sz_{j+1}}=(\1-\ii \sz_j \sz_{j+1})/\sqrt{2}$, and using $h_x=h_z=1$ also $\e^{-\ii \frac{\pi}{4} \sx_j}=(\1-\ii \sx_j)/\sqrt{2}$, and $\e^{-\ii \frac{\pi}{4} \sz_j}=(\1-\ii \sz_j)/\sqrt{2}$. The two 1-qubit gates together perform a permutation of Pauli operators, namely, map $\sx \to \sz, \sy \to \sx, \sz \to \sy$. The two qubit $zz$ gate on the other hand has four $2\times 2$ blocks, in one it maps
\begin{equation}
  \1\sy \to \sz \sx,\qquad \sz\sx \to -\1\sy,
\end{equation}
in the 2nd it maps
\begin{equation}
  \1\sx \to -\sz\sy,\qquad \sz\sy \to \1\sx.
\end{equation}
The remaining two blocks are obtained by swapping the 1st and 2nd qubits (the gate is symmetric with respect to the exchange of qubits). All other Pauli products ($\1\sz,\sz\1,\sx\sy,\sy\sx,\sx\sx,\sy\sy,\sz\sz,\1\1$) are mapped back to themselves, $ab \to ab$. Put all together this causes that at $\tau=\pi/4$ the truncated propagator at $k=0$ has all eigenvalues equal to zero, some of which form Jordan blocks. One has an interesting situation where taking a finite block corresponding to operators with support on $\le r$ sites of an otherwise unitary operator (spectrum on a unit circle) results in a truncated propagator with all eigenvalues being $0$. For instance, for $r=2\times 2$ truncation there is one $2\times2$ Jordan block for each basis class, i.e., there are $10$ (\ref{eq:basis}), in which a product of $\sz$ is mapped to a product of $\sy$ at same positions, for instance,
\begin{equation}
  \begin{smallmatrix}
      \1 & \1 \\
      \sz & \sz
  \end{smallmatrix}
  \to
   \begin{smallmatrix}
      \1 & \1 \\
      \sy & \sy,
   \end{smallmatrix}
\end{equation}
whereas a product of $\sy$ is mapped to an operator with larger support, concretely, $\sy$ are mapped to $\sx$ at the same position, times a product of $\sy$ at all sites neighboring the initial $\sy$ (the above $\sy\sy$ is mapped to a product of two $\sx$ and six new $\sy$ at sites connected to the original $\sy$).

If one has $h_x \neq 1$ things are a bit similar. Many eigenvalues are still zero, but Jordan blocks are destroyed. For $r=2\times2$ one gets exactly one nonzero eigenvalue for each Jordan block. Namely, the eigenvalue corresponding to the class with $p$ Pauli $\sz$ is $\cos^p{(h_x \pi/2)}$, with the corresponding eigenvector having single-site operator $a$,
\begin{equation}
  a=-\sin{(h_x \pi/2)}\sy+\cos{(h_x \pi/2)}\sz,
\end{equation}
at the location of each ``$x$'' in our class description (\ref{eq:basis}). All of the above properties are insensitive to the value of $h_z$. Properties of $M$ at $\tau=\pi/4$ are therefore rather interesting and warrant further study.

\subsection{Special point $\tau=\pi/2$}

Here one has $\e^{-\ii \frac{\pi}{2}\sz_j \sz_{j+1}}=-\ii \sz_j \sz_{j+1}$, $\e^{-\ii \frac{\pi}{2} \sx_j}=-\ii \sx_j$, and $\e^{-\ii \frac{\pi}{2} \sz_j}=-\ii \sz_j$. Because of coordination number $4$ on a 2D square lattice the 2-qubit gates square to one, so that we get $U'_{\pi/2}=\prod_j (\ii \sy_j)$, i.e., a simple product propagator. The two-step propagator is on the other hand trivial $(U'_{\pi/2})^2=(-1)^L \1$, i.e., one has a DTC. This means that all eigenvalues of $U'_{\pi/2}$ are $\pm 1$. This carries over to the spectrum of $M$ which also has all its eigenvalues equal to $\pm 1$. Close to $\tau=\pi/2$ truncations we use, e.g. $3\times 3$, are not large enough to get an accurate estimate for $|\lambda_1|$, for instance, jagged behavior in Fig.~\ref{fig:2d}(A) (for some $\tau$ the largest eigenvalue can even have the modulus larger than $1$).

}


\begin{thebibliography}{99}
\bibitem{foothyper} \new{Hyperbolic systems are, roughly speaking, systems where at each point there is an unstable (stretching) and a stable (folding) manifold. A uniformly hyperbolic system is a one with stretching exponents uniformly larger than $1$, i.e., the stretching never becomes zero, or, stable and unstable directions never become degenerate and coincide.}
\bibitem{marcin} A somewhat similar method has been very recently suggested in Ref.~\cite{marcin25}. For circuits their matrix $K$ is equal to $K=2\cdot\1-M-M^{\rm T}$, where $M$ is our operator propagator at $k=0$.
\bibitem{foot1} Note that we use $J=1/2$ in order to be able to directly compare timescales to e.g. Ref~\cite{Heyl}, which means that the prefactor in front of the $z-z$ term is $\frac{\tau}{4}$, i.e., $\tau=0.8$ can indeed be considered ``small''.
\bibitem{footconv} Convergence is quite fast with $r$, perhaps going as $\sim 1/r^6$ for $r\ge 10$ (or even exponential in $r$). The estimated error bar is conservative. Note that despite rather small gaps $\approx 3\cdot 10^{-5}$ we can with our in-situ propagation method and ARPACK treat large support sizes of $r=13$ where the size of $M$ is $N\approx 5 \cdot 10^7$.
\bibitem{foot2} It would be beneficial to have a set of reference cases (``easy'' and ``hard''), for instance in the kicked Ising model, for which one could compare precise values of diffusion constant in order to assess quality of different numerical methods (Ref.~\cite{white24} is a step in this direction). 
\end{thebibliography}
\end{document}